\documentclass[fleqn,usenatbib]{mnras}
\usepackage{newtxtext,newtxmath}
\usepackage[T1]{fontenc}
\DeclareRobustCommand{\VAN}[3]{#2}
\let\VANthebibliography\thebibliography
\def\thebibliography{\DeclareRobustCommand{\VAN}[3]{##3}\VANthebibliography}
\usepackage{graphicx}
\usepackage{amsmath}	
\usepackage{fancyhdr}
\usepackage{hyperref}
\usepackage{todonotes}
\usepackage{orcidlink}
\usepackage{xcolor}
\usepackage{soul}
\sethlcolor{yellow}

\title[Precision Test of General Relativity with Gravitational Waves]{A Model-Independent Precision Test of General Relativity using Bright Standard Sirens from ongoing and upcoming detectors}

\author[Afroz and Mukherjee]{
  Samsuzzaman Afroz $^{1}$\orcidlink{0009-0004-4459-2981} \thanks{samsuzzaman.afroz@tifr.res.in},
  Suvodip Mukherjee $^{1}$\orcidlink{0000-0002-3373-5236} \thanks{suvodip.mukherjee@tifr.res.in} 
  \\
  $^{1}$Department of Astronomy and Astrophysics, Tata Institute of Fundamental Research, Mumbai 400005, India
}

\date{3 April 2024}
\pubyear{2023}

\begin{document}
\label{firstpage}
\pagerange{\pageref{firstpage}--\pageref{lastpage}}
\maketitle

\begin{abstract}
Gravitational waves (GWs) provide a new avenue to test Einstein’s General Relativity (GR) using the ongoing and upcoming GW detectors by measuring the redshift evolution of the effective Planck mass proposed by several modified theories of gravity. We propose a model-independent, data-driven approach to measure any deviation from GR in the GW propagation effect by combining multi-messenger observations of GW sources accompanied by EM counterparts, commonly known as bright sirens (Binary Neutron Star (BNS) and Neutron Star Black Hole systems (NSBH)). We show that by combining the GW luminosity distance measurements from bright sirens with the Baryon Acoustic Oscillation (BAO) measurements derived from galaxy clustering, and the sound horizon measurements from the Cosmic Microwave Background (CMB), we can make a data-driven reconstruction of deviation of the variation of the effective Planck mass (jointly with the Hubble constant) as a function of cosmic redshift. Using this technique, we achieve a precise measurement of GR with redshift (z) with a precision of approximately 7.9\% for BNSs at redshift z=0.075 and 10\% for NSBHs at redshift z=0.225 with 5 years of observation from LIGO-Virgo-KAGRA network of detectors. Employing Cosmic Explorer and Einstein Telescope for just 1 year yields the best precision of about 1.62\% for BNSs and 2\% for NSBHs at redshift z=0.5 on the evolution of the frictional term, and a similar precision up to z=1. This measurement can discover potential deviation from any kind of model that impacts GW propagation with ongoing and upcoming observations.
\end{abstract}

\begin{keywords}
Gravitational waves, gravitation, cosmology: observations
\end{keywords}

\section{Introduction}
The General Theory of Relativity (GR) has stood as a cornerstone of our understanding of gravity for over a century. It elegantly explains how massive objects warp the fabric of spacetime, causing other objects to move along curved trajectories. However, in extreme astrophysical environments such as black holes and the early universe, the GR faces challenges in its applicability \citep{thorne1995gravitational, sathyaprakash2009physics}.
Recently, a groundbreaking development occurred with the direct detection of gravitational waves (GWs) by the LIGO-Virgo collaboration \citep{abbott2016observation, abbott2016gw151226, scientific2017gw170104, abbott2017gw170817, goldstein2017ordinary}. This achievement has opened a new avenue for testing GR, particularly in the context of compact binary systems existing in relativistic regimes. These systems include Binary Neutron Stars (BNS), Neutron Star-Black Hole pairs (NSBH), and Binary Black Holes (BBH). These compact binaries offer unprecedented opportunities to delve into the fundamental nature of gravity across a vast range of mass scales and cosmological distances. Ground-based detectors like LIGO \citep{aasi2015advanced}, Virgo\citep{acernese2014advanced} and KAGRA \citep{akutsu2021overview} have played pivotal roles in this endeavor. Furthermore, the prospect of future space-based detectors such as LISA \citep{amaro2017laser} and ground-based detectors such as LIGO-Aundha (LIGO-India) \citep{saleem2021science}\footnote{\url{https://dcc.ligo.org/LIGO-M1100296/public}}, Cosmic Explorer(CE) \citep{reitze2019cosmic}, and the Einstein Telescope(ET) \citep{punturo2010einstein} holds even greater promise for advancing our understanding of gravity and rigorously testing GR's predictions. Through these combined efforts, we aim to refine and expand our comprehension of the intricate interplay between gravity and the cosmos. While various models proposing modifications to the theory of gravity have been put forth to address different scenarios, this study is dedicated to testing GR in the propagation of GW through a model-independent, data-driven approach.
GWs and EM signals' propagation speed and luminosity distance offer insight into modified gravity theories. Factors like graviton mass, frictional term(due to running effect of Planck mass) and anisotropic source term contribute to deviations \citep{deffayet2007probing, saltas2014anisotropic, nishizawa2018generalized, belgacem2018gravitational, belgacem2018modified, lombriser2016breaking, lombriser2017challenges}. Precise measurements enable rigorous testing of these alternative theories. Employing bright sirens that emit both GW and EM radiation provides a robust avenue for conducting meticulous assessments of GW propagation speed, graviton mass, and the frictional term($\gamma$(z), see in Equation.\ref{eq:main}). The prompt measurement of the electromagnetic counterpart, occurring approximately 1.7 seconds after the BNS event GW170817, has facilitated the imposition of exceedingly precise constraints on the speed of gravitational wave propagation and the mass of the graviton \citep{abbott2017gw170817, abbott2017gravitational, abbott2019tests}.

\begin{figure*}
\centering
\includegraphics[height=8.3cm, width=18.7cm]{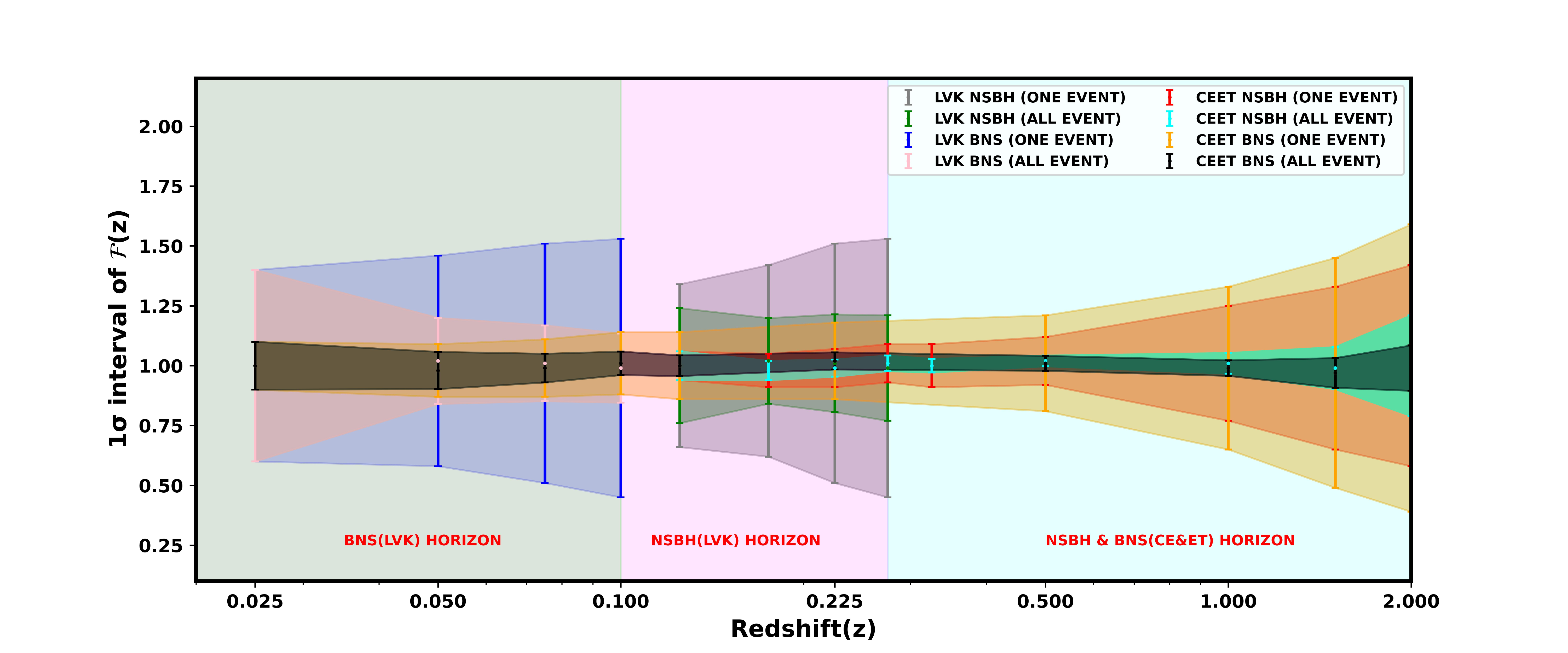}
\caption{This plot showcases the precision achievable in 
reconstructing the redshift variation of the Planck mass captured by $\mathcal{F}(z)$ for both NSBH and BNS systems, employing the current LVK detector with 5 years of observation time and the future CE \& ET detectors with 1 year of observation time. For LVK, the BNS case considers the total events corresponding to a local merger rate $R_0=500$ $\rm Gpc^{-3}yr^{-1}$, while for NSBH, $R_0= 20 \rm Gpc^{-3}yr^{-1}$. For CE \& ET, the BNS scenario assumes $R_0=100$ $\rm Gpc^{-3}yr^{-1}$, and for NSBH, $R_0=20$ $\rm Gpc^{-3}\rm yr^{-1}$ are used. The plot illustrates the enhanced accuracy in measuring the function $\mathcal{F}(z)$ using future detectors compared to our current instruments is possible in the redshift range 0.2 to 1 which is an interesting cosmic epoch when the acceleration of the Universe becomes important.}
\label{fig:fzasz}
\end{figure*}

Various model-dependent parametrizations of $\gamma(z)$ exist in the literature, and one such example is the ($\Xi_0,n$) parametrization \citep{belgacem2018modified} and its detectability is shown for LVK \citep{leyde2022current} and LISA \citep{Baker:2020apq}.

This parameterization depends on two positive parameters, denoted as $\Xi_0$ and n. When $\Xi_0$ is set to 1, it corresponds to GR. There exist several models of modified gravity, each providing predictions for the frictional term in various scenarios. One prominent category is Scalar-Tensor theories. These theories encompass a scalar degree of freedom, the dynamics of which play a pivotal role in the evolution of the universe. Among the most well-known scalar-tensor theories of gravity are the Brans–Dicke theory \citep{brans1961mach}, f(R) gravity \citep{hu2007models} and covariant Galileon models \citep{chow2009galileon}. These belong to the broader class known as Horndeski theories \citep{arai2018generalized}. Beyond Horndeski theories, we have generalized galileon models\citep{gleyzes2014healthy}, Degenerate Higher Order Scalar-Tensor (DHOST) theories \citep{frusciante2019tracker, crisostomi2016extended, achour2016degenerate}. These have been constructed and represent, thus far, the most general scalar-tensor theories that propagate a single scalar degree of freedom alongside the helicity-2 mode of a massless graviton. Additionally, there are targeted scalar-tensor theories specifically designed to remain compatible with observed phenomena \citep{creminelli2017dark, gleyzes2014unifying}. There are several other well-known modified gravity exist in the literature such as f(Q) gravity, f(T) gravity \citep{cai2016f}, bigravity \cite{de2021minimal}, gravity in some extra dimensions \citep{dvali20004d, yamashita2014mapping, corman2021constraining} and so on. In addition, there exist additional parameterizations \citep{belgacem2018modified,romano2022effects}, with the most significant being the polynomial-exponential and exponential parameterizations.

These models exhibit variations from one another\citep{belgacem2019testing} from the power-law form which cannot be captured by $\Xi_0$ and $n$. It is worth mentioning that for this particular parametrization, the fitting formula for $\mathrm{\gamma(z)}$ becomes less accurate in both low and high redshifts in comparison to some of the models (see for example \citep{deffayet2007probing}). Additionally, the intent behind parameterizing them is to facilitate a seamless transition from a unitary value at low redshifts, where the cumulative impacts of modified gravity wave propagation haven't had adequate time to deviations from GR, to a consistent value represented by $\Xi_0$ at higher redshifts. If there is a loss of gravitons into the extra dimensions, it would result in a reduction in the apparent brightness of a GW source. Consequently, we would observe that the luminosity distance for gravitational waves exceeds that of electromagnetic waves \citep{dvali20004d, yamashita2014mapping, corman2021constraining}. In intermediate redshifts, the integrated effect of $\gamma(z)$ may diverge from a simple power-law form in reality. Consequently, it becomes crucial to test GR in a model-independent way and measure the integrated effect of $\gamma(z)$ as a function of the data. To address these issues, we introduce a novel model-independent function, denoted as $\mathcal{F}(z)$, defined in Equation \ref{eq:f(z)}. This function $\mathcal{F}(z)$ is specifically designed to account for any deviation from GR. It serves as a metric for assessing the disparity between the distances denoted $\mathrm{D_{l}^{GW}(z)}$ and $\mathrm{D_l^{EM}(z)}$ so in standard GR this $\mathcal{F}(z)$=1. The model-independent reconstruction technique offers significant advantages by capturing both parametric and non-parametric deviations across any redshift range. This approach efficiently overcomes the constraints encountered in specific models throughout varying redshifts.

Figure \ref{fig:fzasz}, highlighting the enhanced precision in measuring the redshift-dependent variation of the Planck mass ($\mathcal{F}(z)$) for NSBH and BNS systems. Notably, the precision achieved is approximately 7.9\% for BNS systems at redshift 0.075 and 10\% for NSBH systems at redshift 0.225 over a 5-year observation period from the LVK network of detectors. Furthermore, employing CE \& ET for just 1 year yields remarkable precision, reaching about 1.62\% for BNS systems and 2\% for NSBH systems at redshift 0.5, assuming a fixed Hubble constant. However, when considering a varying Hubble constant, the error bar increases to approximately 2 to 2.2 times that of a fixed Hubble constant. Figure \ref{fig:fzasz} also emphasizes that the measurement of $\mathcal{F}(z)$ is particularly improved in the redshift range of 0.2 to 1.0. This improvement is attributed to two main factors: signal-to-noise ratio (SNR) and the number of events. For a more detailed analysis, refer to Section \ref{sec:result}. In the ($\Xi_0$, n) frictional term parametrizations, $\Xi_0$ is set to 1 as the fiducial value. Notably, parameter which controls the redshift evolution denoted by n lacks definition at this point. Measuring n becomes challenging as $\Xi_0$ nears 1 \citep{belgacem2018modified}. Model-independent parametrization enables redshift scaling for F(z), even for the GR value, in contrast to parametric forms.

\begin{figure*}
\centering
\includegraphics[height=8.3cm, width=16.7cm]{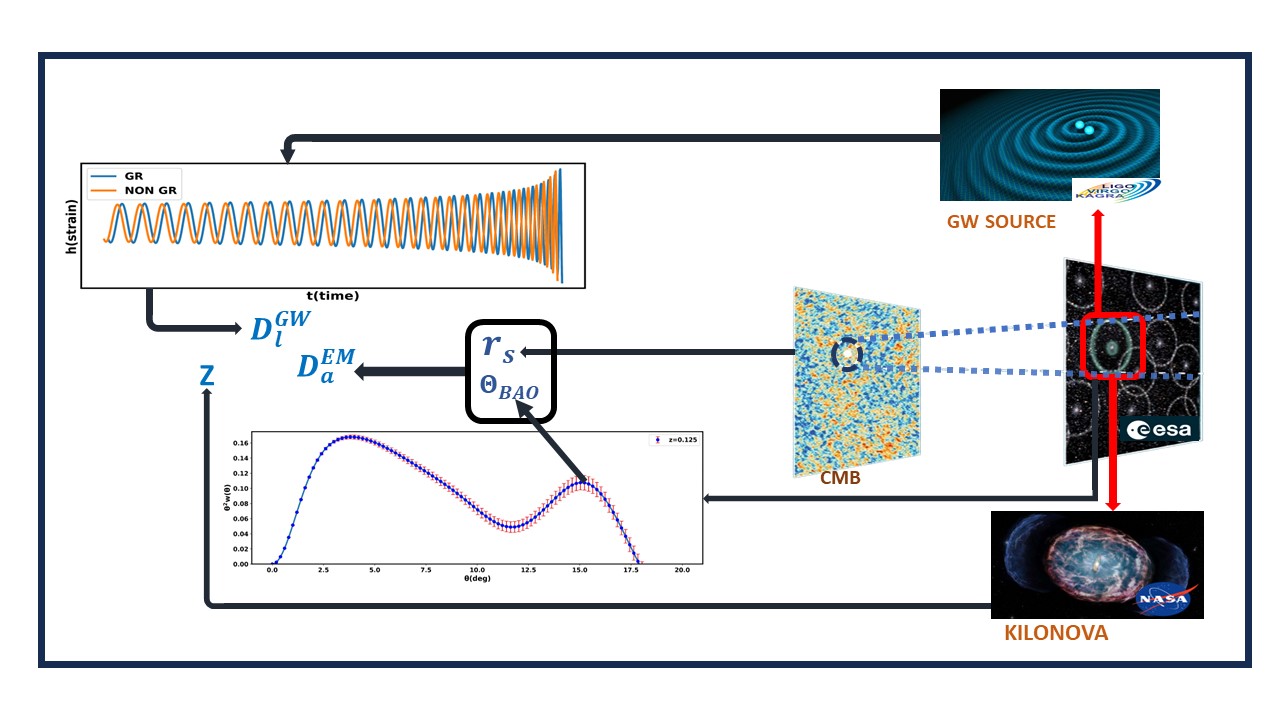}
\caption{This is a schematic diagram demonstrates the measurement of three distinct length scales: the sound horizon ($r_s$), BAO scale ($\theta_{BAO}$), and the GW luminosity distance ($D_l^{GW}$). These measurements are obtained from different independent observables such as  CMB observations, galaxy correlation functions, and gravitational wave sources. The combination of the sound horizon and the BAO scale gives the EM angular diameter distance $D^{\rm EM}_a$, and by using the distance duality relation we can make a data-driven test of GR as a function of redshift using Equation \ref{eq:f}.}
\label{fig:motivation}
\vspace{-0.5cm}
\end{figure*}

The paper is structured as follows: In Section \ref{sec:GWprop}, we delve into the propagation of GW beyond GR. Subsequently, we present our proposed model-independent, data-driven reconstruction of the variation of the effective Planck mass(frictional term). In Section \ref{sec:Mocksamples}, we shift our attention to the astrophysical population of GW sources and explore their detection using both current and future detectors. Section \ref{sec:BAO} is dedicated to a comprehensive discussion on the model-independent measurement of the BAO scale from the galaxy power spectrum.
In Section \ref{sec:Forecast}, we provide a detailed exposition of the formalism underpinning our work. Moving forward, Section \ref{sec:result} encapsulates the main findings of our methodology, while Section \ref{sec:conclusion} serves as the culmination, summarizing our key discoveries and engaging in a discourse on future prospects.

\section{Non-parametric reconstruction of deviation from General Relativity: Formalism}
\label{sec:GWprop}

The propagation of GW in the fabric of spacetime with speed $c$, as described by GR, can be mathematically expressed as follows:
\begin{equation}
\mathrm{h_{(+,\times)}^{''} + 2\mathcal{H}h_{(+,\times)}^{'} + c^2k^2h_{(+,\times)} = 0},
\end{equation}
where $h_{(+,\times)}$ represents the GW strain for plus ($+$) and cross ($\times$) polarizations, the prime symbol denotes the derivative with respect to the conformal time ($\eta$), and $\mathcal{H}$ represents the Hubble parameter in comoving coordinates. This equation serves as a foundational framework for examining the propagation of GWs and evaluating the validity of GR.

In the formulation of modified theories of gravity, several additional parameters come into play. Firstly, there's the frictional term denoted by $\gamma(z)$, representing the influence of friction in the theory. Additionally, parameters such as the speed of GW propagation, the graviton mass, and the anisotropic stress term are considered. Within the framework of GR, all these additional parameters typically assume a fiducial value of zero, except for the speed of gravity, which corresponds to the speed of light ($c$). Therefore, in the context of testing GR through propagation, it's essential to scrutinize these additional parameters to ascertain whether they deviate from the values predicted by GR.

Recent measurement from GW170817 event, have provided stringent constraints indicating that the graviton mass is zero, and thus the speed of GW propagation is equivalent to the speed of light ($c$)\citep{abbott2017gravitational, abbott2019tests}. Modifications to the anisotropic stress term affect the phase of binary waveforms. Recent observations of BBH mergers, such as GW150914 and GW151226, have set some limits on such modifications, although they are not currently very stringent \cite{abbott2016binary}.

As a consequence, $\gamma(z)$ becomes the primary parameter of interest for our study. Consequently, the propagation of GW within the framework of the modified theory of gravity of our interest is described by the equation
\begin{equation}
\mathrm{h_{(+,\times)}^{''} + 2(1-\gamma(z))\mathcal{H}h_{(+,\times)}^{'} + c^2k^2h_{(+,\times)} = 0}.
\label{eq:main}
\end{equation}
The presence of the extra frictional term $\mathrm{\gamma(z)}$ introduces a notable alteration in the amplitude of the GW signal received from a source situated at cosmological distances. This distinction is particularly intriguing as it implies that the luminosity distance measured with standard sirens deviates in principle from that obtained using standard candles or other electromagnetic probes such as CMB or BAO\citep{amendola2018direct,matos2023model}. This distinction arises due to the incorporation of the frictional term within the framework of modified gravity theory. In contrast, standard electromagnetic theory assumes no such frictional term, ensuring that the luminosity distance measured with standard sirens remains unaffected as redshift changes.

This discrepancy presents a compelling opportunity for the detection of modified gravity, potentially serving as a "smoking gun" signature of such modifications. So as a consequence the GW luminosity distance $\mathrm{D_l^{GW}}$ is related to the electromagnetic wave-based luminosity distance $\mathrm{D_l^{EM}}$ by an exponential factor, where the exponent involves integrating $\mathrm{\gamma(z')/(1+z')}$ with respect to $z'$ \citep{belgacem2018gravitational}. Consequently, the expression becomes

\begin{equation}
\mathrm{D_{l}^{GW}(z) = \exp\left(-\int dz'\frac{\gamma(z')}{1+z'}\right) D_l^{EM}(z)}.
\label{eq:gamma_z}
\end{equation}

In our investigation, we focus on tensor perturbations, where the impact is encoded in the non-trivial function $\mathrm{D_{l}^{GW}(z)/D_l^{EM}(z)}$. In this study, we propose a non-parametric reconstruction of the deviation from GR as

\begin{equation}
\mathrm{D_{l}^{GW}(z)={\mathcal{F}}(z)D_l^{EM}(z)}.
\label{eq:f(z)}
\end{equation}
The term ${\mathcal{F}}(z)$ can capture any deviation from GR as a function of redshift. The reconstruction of this quantity from observation of $\rm{D_{l}}^{GW}(z)$ and $\rm{D_{l}}^{EM}(z)$ can capture any modified gravity models which predict a modification in the GW propagation \citep{belgacem2019testing}. Alternatively, a parametric form using $\Xi_0$ and $n$, $\rm{D_{l}^{GW}(z)/D_l^{EM}(z)=\Xi_0+\frac{1-\Xi_0}{(1+z)^n}}$ is also used  to capture all those models which predicts a power-law modification with redshift of the GW propagation effect \citep{belgacem2018modified}.  

In the realm of binary systems, GW offers a unique method for determining luminosity distance, denoted as $\mathrm{D_{l}^{GW}}$. This distance measurement is derived from GW events. While the luminosity distance can also be determined using the standard $\Lambda$CDM cosmology, such an approach is inherently model-dependent. To obtain a model-independent data-driven inference of luminosity distance, we utilize EM luminosity distance calculations derived from various length scales. 

\textit{Data-driven inference of EM luminosity distance:} A data-driven approach to infer the EM luminosity distance can be made from the angular diameter distance $D^{\rm EM}_A(z)$ measured using BAO angular peak position ($\theta_{\rm BAO}(z)$) from large scale structure observation  at different redshifts \citep{peebles1973statistical}, and using the distance duality relation which connects the angular diameter distance with the EM luminosity distance by $D^{\rm EM}_a(z)= D^{\rm EM}_l(z)/(1+z)^2$. The position of the BAO peak in the galaxy two-point correlation function can be inferred directly from observation. The angular scales are related to the comoving sound horizon until the redshift of drag epoch ($z_d \sim 1020$) as 
\begin{equation}
r_s=\int_{z_d}^{\infty}\frac{dz\,c_s(z)}{H_0\sqrt{\Omega_m(1+z)^3+\Omega_\Lambda + \Omega_r(1+z)^4}}.
\end{equation}
By replacing $D_l^{\rm EM}$ in Equation \eqref{eq:f(z)} with the $\theta_{\rm BAO}(z)$ and by using the distance duality relation for EM probes, one can write the above equation as \citep{mukherjee2021testing}
\begin{equation}\label{eq:f}
\mathcal{F}\mathrm{(z)=\frac{D_l^{GW}(z)\theta_{BAO}(z)}{(1+z)r_s}}.
\end{equation}
The comoving sound horizon $r_s$ at the drag epoch can be related to the comoving sound horizon $r_*$ at the redshift of recombination ($z_*= 1090$) by $r_d \approx 1.02r_*$. The quantity $r_*$ is inferred from CMB observation, using the position of the first peak in the CMB temperature power spectrum $\theta_*= r_*/(1+z_*)D^{EM}_a(z_*)$ \citep{spergel2007three, komatsu2011seven, hinshaw2013nine, aghanim2020planck}.

In the above equation, $D_l^{GW}(z)$ is measured from GW sources, $\theta_{\rm BAO}$ is measured from galaxy two-point correlation function, and $r_s$ is inferred from the CMB temperature fluctuations. The redshift to the GW source can be inferred from the EM counterpart for a bright stand siren. As a result, the product of all these observable quantities will lead to an identity value at all redshifts if GR and the fiducial $\Lambda$CDM model of cosmology is the correct theory. However, if there is any departure from these models, then we can measure a deviation from one with redshift. It is important to note that the value of $r_s$ depends on the redshift above $z_d=1020$. Though it depends on the model of cosmology, it is a redshift-independent number of a constant value for the low redshifts $z<2$. So, even if there is an inaccuracy in the inference of true $r_s$, the value of $\mathcal{F}(z)$ will vary by an overall normalization. But the reconstructed redshift evolution of $\mathcal{F}(z)$ will remain the same. In one of the later sections, we will show results for both cases, (i) only $\mathcal{F}(z)$ and (ii) $\mathcal{F}(z)$ and $H_0$ together. The second case will make it possible to marginalize over any inaccurate inference of sound horizon $r_s$  due to incorrect inference of the Hubble constant.

Several alternative models of GR predict deviation in different natures of the acceleration of the Universe that are predicted from the $\Lambda$CDM. As a result, this data-driven approach can make it possible to explore both deviation from GR and $w=-1$ equation of state of dark energy. The use of different observational probes was done previously \citep{mukherjee2021testing} for a parametric form of deviation from GR (using $\Xi_0$ and $n$) valid for a class of model. However, the non-parametric form using $\mathcal{F}\mathrm{(z)}$ proposed in this analysis, can make a redshift-dependent reconstruction of any deviation from GR using the multi-messenger observations.

The redshift can be deduced through the EM counterparts associated with these bright sirens. Ongoing and upcoming missions dedicated to identifying EM counterparts include DESI \citep{collaboration2023early}, Vera Rubin \citep{abell2009lsst}, Fermi \citep{bissaldi2009ground}, Swift \citep{burrows2005swift}, HST \citep{scoville2007cosmos}, Roman space telescope \citep{eifler2021cosmology}, Chandra \citep{vikhlinin2009chandra}, VLA \citep{benz1989vla}, ZTF \citep{bellm2014zwicky}, Astrosat \citep{agrawal2006broad}. A comprehensive article on the required multi-messenger observations can be found here \citep{Ronchini:2022gwk}. For the measurement of BAO scale, ongoing and upcoming missions are committed to spectroscopic surveys. Notable projects in this regard include eBOSS  \citep{alam2017clustering}, DESI \citep{collaboration2023early}, Vera Rubin \citep{abell2009lsst}, Euclid \citep{laureijs2011euclid}, and SPHEREx \citep{dore2018science}. CMB measurements from the ongoing and upcoming missions include ACTPol \citep{thornton2016atacama}, Simons Observatory \citep{abitbol2019simons}, SPT-3G \citep{sobrin2022design} and CMB-S4 \citep{abazajian2016cmb}.

\section{GW mock samples for LVK, Cosmic Explorer, and Einstein Telescope}
\label{sec:Mocksamples}
\subsection{Modelling of the astrophysical population of GW source}
To accurately assess deviations from the GR model, it is crucial to rely on real GW events. The total number of GW events is contingent upon the merger rates, and mass population of the GW sources. We will explain below the astrophysical models used in the analysis. 

\texttt{Merger rate: } In this study we use a delay time model of the binary mergers  \citep{ochsner2012richard, dominik2015double}. In the delay time model, the merger rate is described in terms of the delay time distribution, denoted as $t_d$. The delay time refers to the elapsed time between the formation of stars that will eventually become black holes and the actual merging of these black holes. It is important to note that the time delay is not uniform across all binary black holes but instead follows a specific distribution. This distribution function accounts for the variations in the delay time and is defined as follows:

\begin{equation}
    p_t(t_d|t_d^{min},t_d^{max},d) \propto 
    \begin{cases}
    (t_d)^{-d} & \text{, for }  t_d^{min}<t_d<t_d^{max}, \\
    0 & \text{otherwise}
    \end{cases}
\end{equation}

The delay time is given by $t_d=t_m-t_f$, where $t_m$ and $t_f$ are the lookback times of merger and formation respectively \citep{karathanasis2023binary}. So the merger rate at redshift z can be defined as

\begin{equation}
    \mathrm{R_{TD}(z)=R_0\frac{\int_z^{\infty}p_t(t_d|t_d^{min},t_d^{max},d)R_{SFR}(z_f)\frac{dt}{dz_f}dz_f}{\int_0^{\infty}p_t(t_d|t_d^{min},t_d^{max},d)R_{SFR}(z_f)\frac{dt}{dz_f}dz_f}},
\end{equation}

The parameter $R_0$ represents the local merger rate, indicating the frequency of mergers at a redshift of z=0. According to the study in \citet{abbott2023population}, the estimated values of $R_0$ for BNS systems vary between 10 $\rm Gpc^{-3} yr^{-1}$ and 1700 $\rm Gpc^{-3}yr^{-1}$. For NSBH systems, the $R_0$ values range from 7.8 $\rm Gpc^{-3} yr^{-1}$ to 140 $\rm Gpc^{-3}yr^{-1}$.

{In this study, our analysis centers exclusively on the propagation of gravitational waves, employing the delay time merger rate model to sample binary systems. The considered modified theory of gravity does not influence binary system formation, ensuring the merger rates remain unaffected.}

In our approach, we adopt a standard local merger rate of $R_0$ = 20 $\rm Gpc^{-3} yr^{-1}$ for NSBH systems. For BNS systems, we consider four different scenarios with $R_0$ values of 100, 200, 300, and 500 $\rm Gpc^{-3} yr^{-1}$ respectively. This methodology allows us to examine the effects of different local merger rates on our results.

The numerator of the expression involves the integration over redshift $z_f$ from $z$ to infinity, where $p_t(t_d|t_d^{min},t_d^{max},d)$ is the delay time distribution, $R_{SFR}(z_f)$ is the star formation rate, and $\frac{dt}{dz_f}$ is the jacobian of the transformation. The star formation rate($R_{SFR}(z)$), is determined using the \citet{madau2014cosmic} star formation rate.

The total number of compact binary coalescing events per unit redshift is estimated as 
\begin{equation}
\mathrm{\frac{dN_{GW}}{dz} = \frac{R_{\rm TD}(z)}{1+z} \frac{dV_c}{dz} T_{obs}},
\label{eq:totno}
\end{equation}
where $T_{obs}$ indicates the total observation time, $\frac{dV_c}{dz}$ corresponds to the comoving volume element, and $R(z)$ denotes the merger rate \citep{karathanasis2022gwsim}. We consider the delay time merger rate with a specific delay time $t_d =500\text{Myrs}$ and a power-law exponent of $d=1$. However, it is regrettable to note that not all of these events can be detected due to the limitations imposed by the sensitivity of our current detectors. To determine which events are detectable, the calculation of the matched filtering signal-to-noise ratio (SNR) plays a crucial role. The SNR serves as a measure of the strength of the gravitational wave signal relative to the background noise. Only those events with a matched filtering SNR greater than or equal to a predetermined threshold SNR ($\rho_{\rm TH}$) can be reliably detected. \citep{maggiore2007gravitational}.  

For a GW emitted by an optimally oriented binary system, the optimized SNR, denoted as $\rho$, is defined as follows \citep{sathyaprakash1991choice, cutler1994gravitational, balasubramanian1996gravitational, nissanke2010exploring}

\begin{equation}
{\rho^2 \equiv 4\int_{f_{\text{min}}}^{f_{\text{max}}} df \frac{|h(f)|^2}{S_n(f)}},
\label{snr}
\end{equation}

Here, $S_n(f)$ represents the power spectral density of the detector. The function $h(f)$ corresponds to the Gravitational Wave (GW) strain in the restricted post-Newtonian approximation and is defined for plus ($+$) and cross ($\times$) polarization as \citep{ajith2008template}:
\begin{equation}
    \mathrm{h(f)_{\{+, \times\}}=\sqrt{\frac{5\eta}{24}}\frac{(GM_c)^{5/6}}{D_l\pi^{2/3}c^{3/2}}}f^{-7/6}e^{\iota\Psi(f)}\mathcal{I}_{\{+, \times\}}.
\end{equation}

In this expression, the symbol $\eta$ represents the symmetric mass ratio. The term $M_c$ signifies the chirp mass of the system. The variable $D_l$ denotes the luminosity distance. The constant c represents the speed of light in a vacuum. $\mathcal{I}_{+}= (1+\cos^{2}i)/2$ and $\mathcal{I}_{\times}= \cos i$ depends on the inclination angle $i$. Finally, $\Psi(f)$ stands for the phase of the waveform.

For the inspiral phase, this formula has proven to be accurate, particularly given the moderate mass ratio in our case. We employed this waveform model exclusively for SNR calculations. Given the moderate mass ratio, the impact of higher-order modes on SNR is minimal. Since SNR is tied to the number of events, the inclusion of higher-order Post-Newtonian corrections does not affect the individual events' error on $\mathcal{F}(z).$ The parameter estimation for the selected sources, conducted with \texttt{Bilby} \citep{ashton2019bilby}, utilized the \texttt{IMRPhenomHM} waveform model \citep{kalaghatgi2020parameter}.

However, the signal detected by a GW detector h$_{\rm det}$ is a complex interplay of several variables, including the detection antenna functions ($F_{+}, F_{\times}$), and can be expressed as:

\begin{equation}
    h_{\rm det}=F_{+}h_{+}+F_{\times}h_{\times},
\end{equation}
here $F_{+}$ and $F_{\times}$ are the antenna functions defined as follows \citep{finn1993observing}
\begin{align}\label{eq:antenna}
    &\mathrm{F_{+}=\frac{1}{2}(1+cos^2\theta)cos2\phi cos2\psi-cos\theta sin2\phi sin2\psi},\\\nonumber
     &\mathrm{F_{\times}=\frac{1}{2}(1+\cos^2\theta)\cos2\phi sin2\psi+cos\theta sin2\phi cos2\psi},
\end{align}
where $\theta$ and $\phi$ define the location of the source in the sky, and $\psi$ is related to the orientation of the binary system with respect to the detector. 
Consequently, the matched filtering SNR ($\rho$) takes the form \citep{finn1996binary}
\begin{equation}
    \rho = \frac{\Theta}{4}\biggl[4\int_{f_{min}}^{f_{max}}h(f)^2/S_n(f)df\biggr]^{1/2},
    \label{eq:SNR}
\end{equation}
where $\Theta^2 \equiv 4 \left(F_{+}^2(1+\cos^2i)^2 + 4F_{\times}^2\cos^2i\right)$. Averaging over many binaries inclination angle and sky positions, \citet{finn1996binary} showed that $\Theta$ follows a distribution 
\[ P_{\Theta}(\Theta) = 
\begin{cases}
5\Theta(4-\Theta)^3/256& \text{if } 0<\Theta<4,\\
0,              & \text{otherwise}.
\end{cases}\]

\texttt{Mass model:} The mass population used for BNS and NSBHs are motivated by the recent results from the third catalog of GW sources by the LVK collaboration \citep{abbott2023population, talbot2018measuring, abbott2019binary}. To model the primary mass distribution for black holes, we adopt a Power Law + Gaussian Peak model with smoothing, which combines a power law distribution with the inclusion of a Gaussian component encompassing masses ranging from $10M_{\odot}$ to $20M_{\odot}$. Conversely, we assume a uniform distribution of masses for the secondary component of the neutron star, falling within the range of $1M_{\odot}$ to $2M_{\odot}$.

In this study, we employ two sets of GW detectors: one comprising LIGO \citep{aasi2015advanced}, Virgo\citep{acernese2014advanced} and KAGRA \citep{akutsu2021overview}(LVK), and the other consisting of Cosmic Explorer \citep{reitze2019cosmic} and Einstein Telescope \citep{punturo2010einstein} (CE \& ET). 

As we have primarily focused on BNS and NSBHs, we did not include LISA in our analysis. Despite its ability to reach deeper into the redshift and provide more precise parameter measurements, the range of masses of compact objects considered in our analysis is comparatively lighter than the masses detectable from LISA. The LISA bright siren analysis will be presented in a future work.

For the LVK system, a threshold SNR of $\rho_{\rm Th} =12$ is chosen. For CE \& ET, the threshold SNR is adjusted to enable the detection of events up to a redshift of 3 (which corresponds to an SNR of 25 and 55 for BNS and NSBH respectively.). A redshift bin size of $\Delta z=0.025$ is adopted, and within this bin, the total number of events is computed using Equation \ref{eq:totno}. Events are generated based on distances inferred from redshifts, the parameter $\Theta$, and object masses from their respective distributions.

For NSBH systems, the LVK's 5-year operation is projected to yield approximately 30 detectable events (all of them need not have a detectable EM counterpart). However, with a one-year observation period, the CE \& ET are expected to identify around 600 events. In the case of BNS systems, the LVK's 5-year operation is anticipated to result in approximately 25, 39, and 60 detectable events for local merger rates ($R_0$) of 200, 300, and 500 $\rm Gpc^{-3} yr^{-1}$ respectively. On the other hand, CE \& ET are expected to detect approximately 4800 events in a one-year observation period with $R_0$=100.

\begin{figure}
\centering
\includegraphics[height=6.5cm, width=9cm]{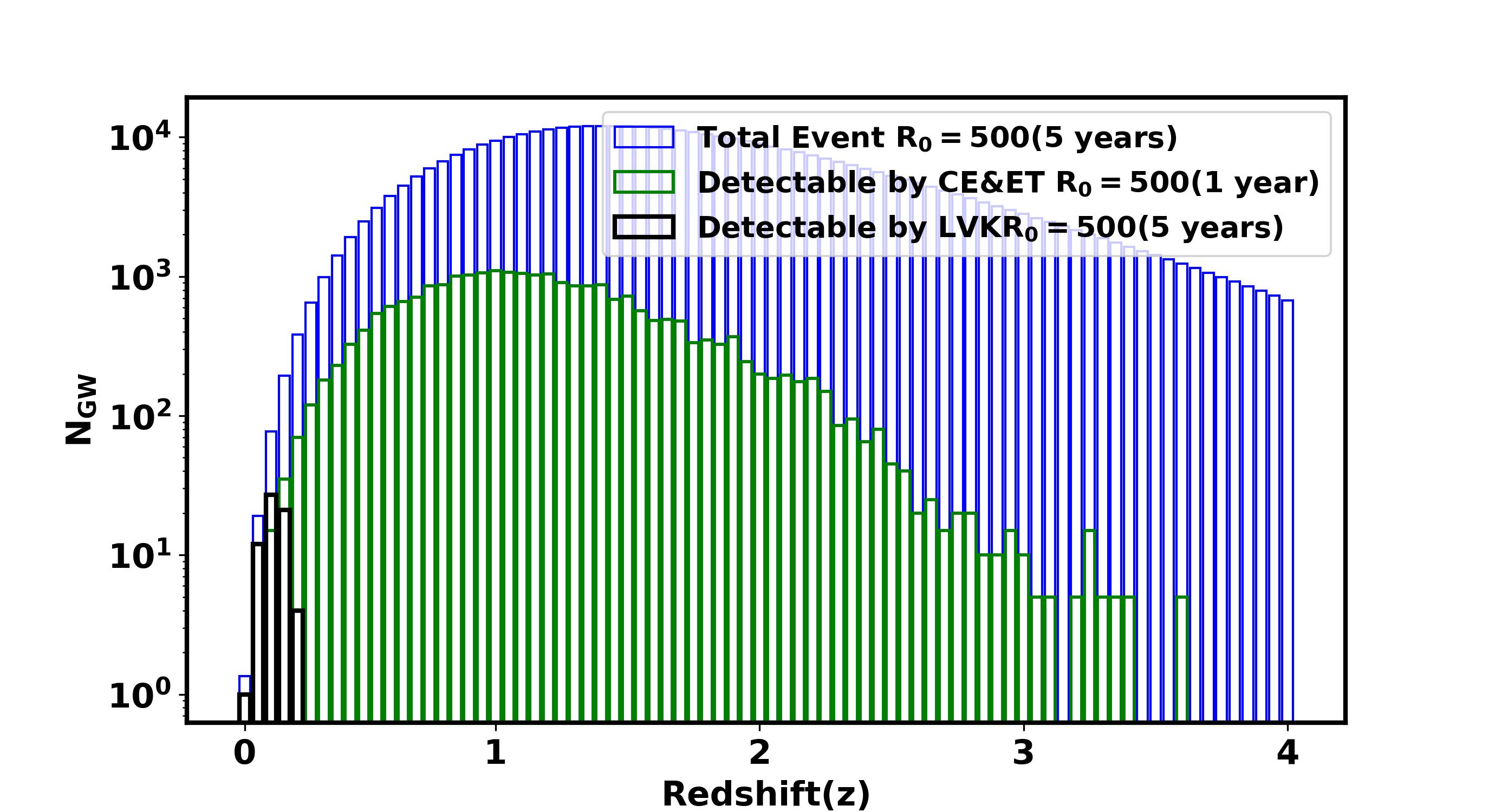}
\caption{Total number of injected events (in blue) and the number of detectable GW events as a function of redshift for BNS sources are shown for CE \& ET (in green) and LVK (in black). The total observation periods are mentioned in parentheses.}
\label{fig:BNSevent}
\end{figure}

\begin{figure}
\centering
\includegraphics[height =6.5cm, width=9cm]{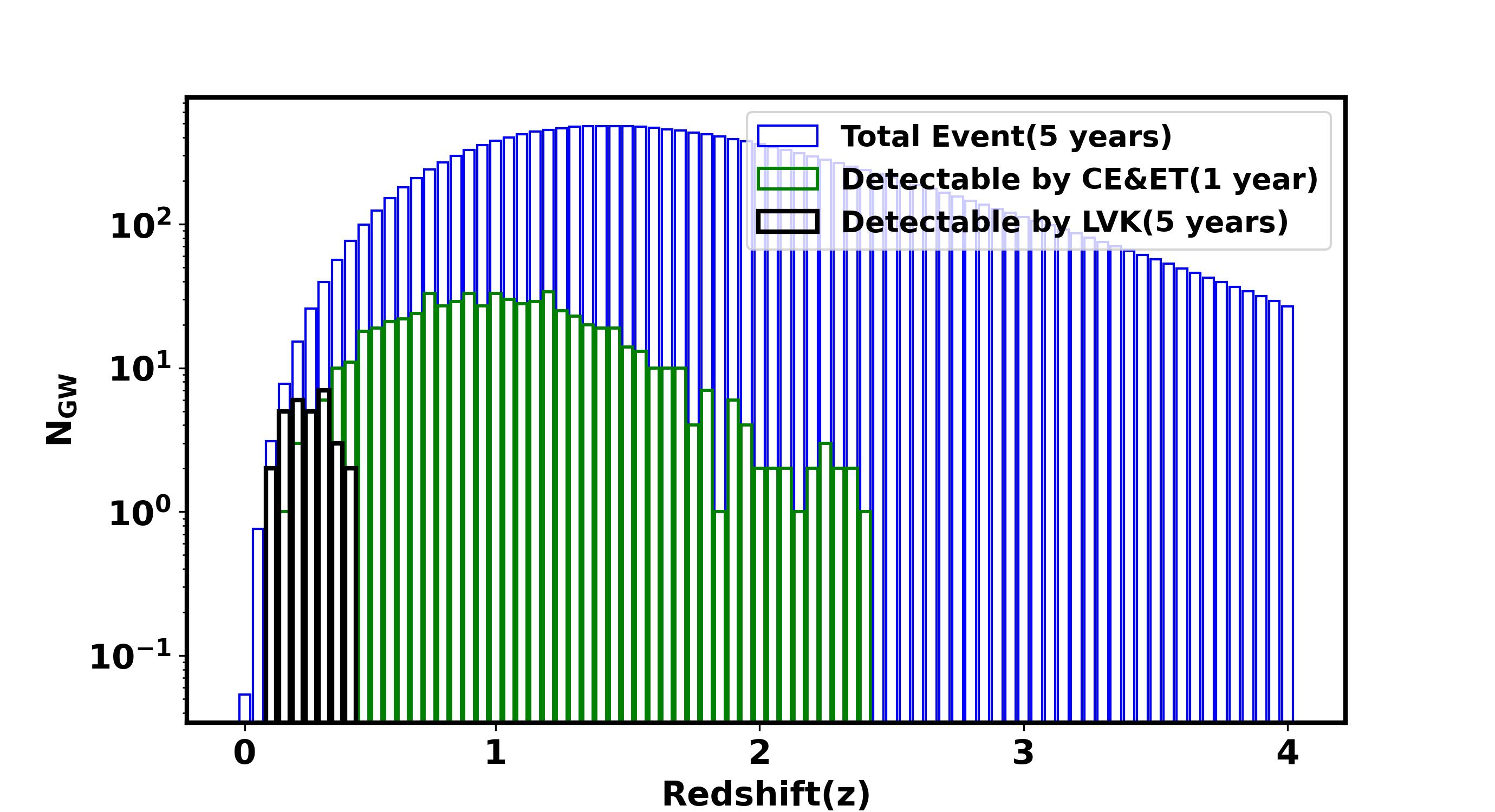}
\caption{Total number of injected events (in blue) and the number of detectable GW events as a function of redshift for NSBH sources are shown for CE \& ET (in green) and LVK (in black). The total observation periods are mentioned in parentheses.}
\label{fig:NSBhevent}
\end{figure}

In Figure \ref{fig:BNSevent} and Figure \ref{fig:NSBhevent}, we present the projected total and detectable coalescing events involving BNS and NSBH systems within specific redshift intervals ($\Delta z$). To obtain these estimates, we first determine the number of merging events for each redshift bin using Equation \ref{eq:totno}. For each event, we employ the inverse transform method to derive the primary mass, secondary mass, and $\Theta$ from their respective probability distributions. The primary mass is generated within the range of $10M_{\odot}$ to $20M_{\odot}$, the secondary mass within $1M_{\odot}$ to $2M_{\odot}$, and $\Theta$ within the interval 0 to 4. 

Incorporating redshift information for each event, we calculate the corresponding luminosity distance and SNR using Equation \ref{eq:SNR} for a single detector, which takes into account the orientation and inclination angle. This approach enhances the realism of our analysis by considering not only optimally oriented binary systems but also accounting for various orientations and inclination angles, providing a more comprehensive understanding of the system dynamics. However, the parameter estimation of the sources is performed using \texttt{Bilby}.

The total SNR denoted as $\rho_{\rm total}$, is then determined by combining individual SNRs from individual detectors using the formula $\rho_{\rm total}=\sqrt{\sum_i \rho_i^2}$. This detection capability is evaluated using both the CE \& ET and LVK network of GW detectors over a defined observation period, $T_{OBS}$. The depicted curve in the graph illustrates the distribution of these events up to a redshift of z = 4. This representation offers valuable insights into the temporal occurrence of these mergers throughout cosmic history and the likelihood of their successful detection using current and future planned instrumentation.

\subsection{GW Source Parameter Estimation using Bilby} We initiate the parameter estimation process for these discernible sources by employing the \texttt{Bilby} \citep{ashton2019bilby} package, which yields realistic posterior distributions on the GW luminosity distance marginalized over the other source parameters. The mass of the GW sources and the number of sources with redshift are drawn as described in the last section. For the remaining source parameters, such as the inclination angle ($i$), polarization angle ($\psi$), GW phase ($\phi$), right ascension (RA), and declination (Dec), we sample uniformly, considering a non-spinning system. With these injected parameters, we proceed to generate a GW signal using the \texttt{IMRPhenomHM} \citep{kalaghatgi2020parameter} waveform model. This model incorporates higher-order modes that can alleviate the degeneracy between the luminosity distance($D_l$) and inclination angle($i$) for unequal mass sources. By constraining the priors of all other parameters to delta functions \footnote{We have assumed here that the RA and Dec are inferred from the EM counterpart.}, we generate posterior distributions for $m_1$, $m_2$, $D_l$ and $i$. Among these, the posterior of $D_l$ is particularly crucial for our study, as it will be used for the inference of $\mathcal{F}(z)$. Additionally, the detection probability of the electromagnetic counterpart is contingent on the value of the inclination $i$ \citep{chase2022kilonova, Ronchini:2022gwk}.

\section{Baryon Acoustic Oscillation Scale from galaxy power spectrum}
\label{sec:BAO}
\texttt{Inference of BAO from galaxy two-point angular correlation function:} The BAO arises from the intricate interplay between radiation pressure and gravity during the early stages of the Universe. Previous works on this and corresponding measurements can be found here \citep{peebles1973statistical, crocce2011modelling}.  As photons and baryons decouple, a specific length known as the sound horizon at the drag epoch ($z_d$), represented by $r_s$, leaves a distinct signature on the distribution of galaxies and the power spectrum of CMB anisotropies \citep{peebles1970primeval, sunyaev1970small, bond1987statistics, hu2002cosmic, blake2003probing}. By examining the two-point angular correlation function (2PACF) of galaxy catalogs denoted by $w(\theta)$, one can identify the BAO signature, which manifests as a prominent feature resembling a bump. The BAO scale can be derived by fitting the angular correlation function $w(\theta)$ with a model defined as a combination of a power-law and a Gaussian component as \citep{xu20122, carvalho2016baryon, alam2017clustering}
 
\begin{equation} \mathrm{w(\theta, z)=a_1+a_2\theta^k+a_3exp\biggl(-\frac{(\theta-\theta_{FIT}(z))^2}{\sigma_{\theta}^2}\biggr)},
\label{eq:fit}
\end{equation}
where, the coefficients $a_i$ represent the parameters of the model, $\sigma_{\theta}$ denotes the width of the BAO feature, and $\theta_{FIT}$ corresponds to the best-fit value of the angular BAO scale. This signature serves as a robust standard ruler, enabling independent measurements of the angular diameter distance denoted as $D_a(z)$ \citep{peebles1974statistical, davis1983survey, hewett1982estimation, hamilton1993toward, landy1993bias}. 

\texttt{Modelling of the 2PACF:} Theoretically, one can model the 2PACF $w(\theta)$  as 
\begin{equation}
    \mathrm{w(\theta)=\sum_{l\geq 0}\biggl(\frac{2l+1}{4\pi}\biggr)P_l(cos(\theta))C_l},
\end{equation}
where $P_l$ is the Legendre polynomial of the $l^{th}$ order and $C_l$ the angular power spectrum, which can be expressed in terms of the three-dimensional matter power spectrum $\mathcal{P}(k)$ as follows \citep{peebles1973statistical, crocce2011modelling}
\begin{equation}
\mathrm{C_l=\frac{1}{2\pi^2}\int4\pi k^2\mathcal{P}(k)\psi_l^2(k)e^{-k^2/k_{eff}^2}},
\end{equation}
where the quantity $\psi_l(k)$ is defined as:  
\begin{equation}
    \mathrm{\psi_l(k)=\int dz\phi(z)j_l(kr(z))},
\end{equation}
where $\phi(z)$ is the galaxy selection function and $j_l$ is the spherical Bessel function of the $\mathrm{l^{th}}$ order, and r(z) is the comoving distance.
To improve the convergence of the integral for the angular power spectrum, a damping factor $e^{(-k^2/k_{eff}^2)}$ is introduced \citep{anderson2012clustering}. Throughout the calculations, we adopt a fixed value of $1/k_{eff}$ = 3 Mpc/h, which has no significant impact on the angular power spectrum for the scales of interest. 

\begin{figure}
\centering
\includegraphics[height=6.5cm, width=9cm]{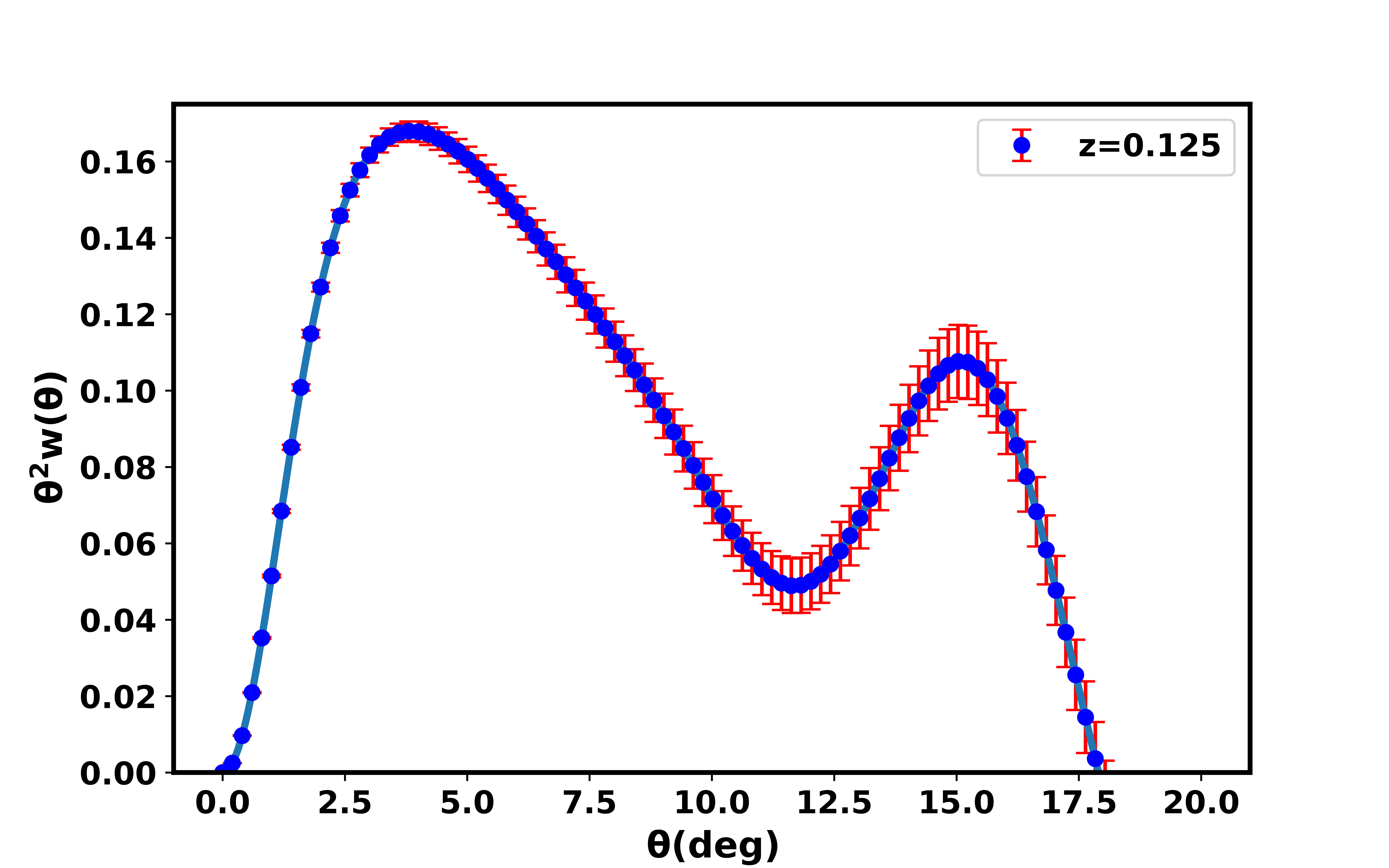}
\caption{This plot illustrates the relationship between the 2PACF($\theta^2w(\theta)$) is shown as a function of $\theta$ for a specific redshift, z=0.125 calculated using the linear matter power spectrum. Notably, a distinct bump is observed around $\theta \sim 16.0$, indicating the presence of the BAO characteristic scale $\theta_{BAO}$.}
\label{fig:bao}
\vspace{-0.5cm}
\end{figure}

In this study, we calculate the matter power spectrum from the module \texttt{CAMB} \citep{lewis2011camb}, and the selection function on redshift is formulated as a normalized Gaussian function. The standard deviation of this Gaussian function is set to be 5\% of the mean value. The anticipated photo-z error for the Vera Rubin Observatory is described by the expression $0.03(1+z)$, as referenced in \citep{mandelbaum2018lsst}. Figure \ref{fig:bao} visually demonstrates how the product of $\theta^2w(\theta)$ changes with angular separation at a specific redshift $z=0.125$. The prominent peak at $\theta \sim 16.0$ is a significant observation, highlighting the characteristic scale associated with BAO, which is crucial for our study.

\begin{figure}
\centering
\includegraphics[height=6.5cm, width=10cm]{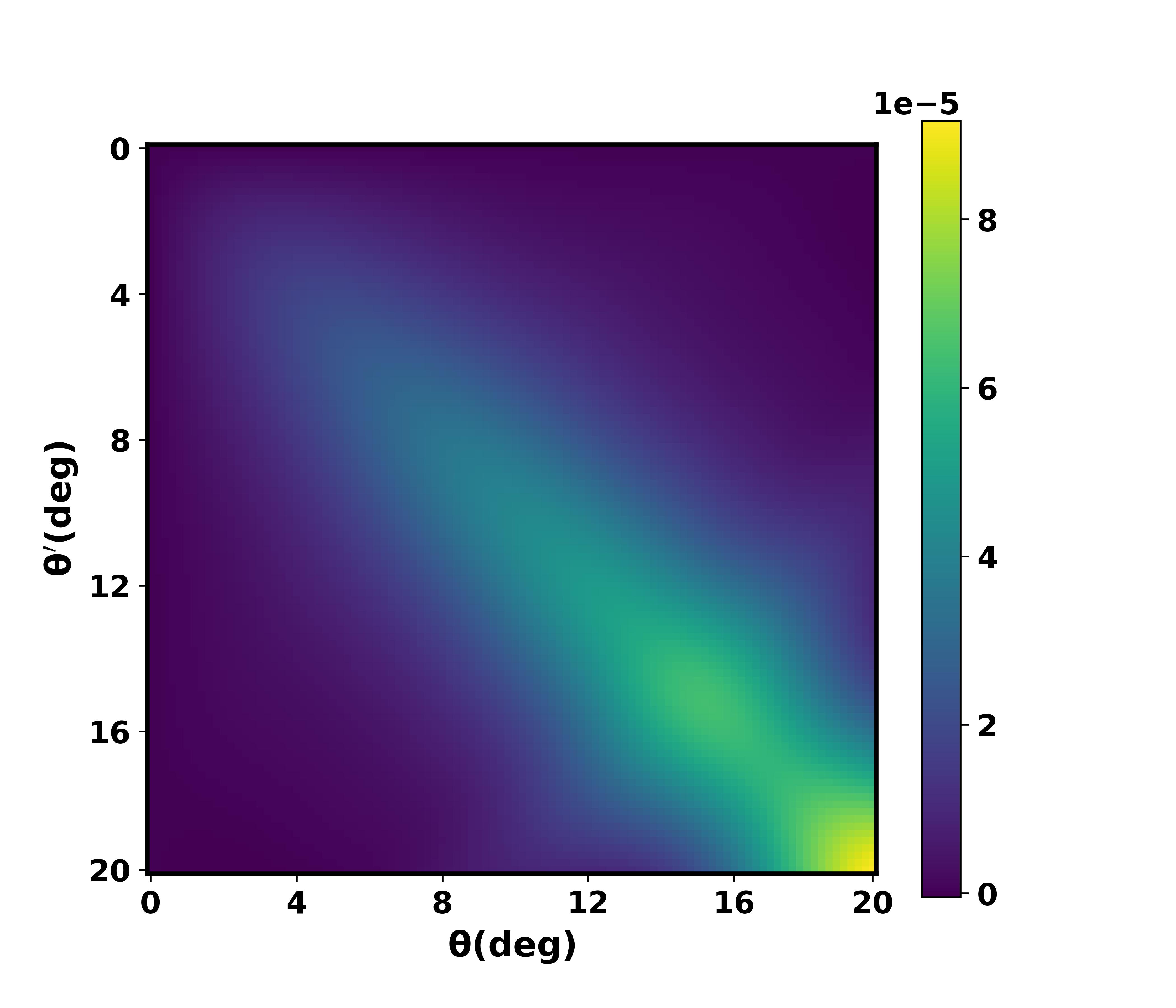}
\caption{This plot displays the covariance matrix($\mathrm{Cov_{\theta\theta'}}$) calculated using linear matter power spectrum as a function of $\theta$ and $\theta'$ for a specific redshift, z=0.125.  It is evident that as both $\theta$ and $\theta'$ increase, the value of the covariance matrix also increases indicating larger error at the large angular scales.}
\label{fig:covmatplot}
\vspace{-0.5cm}
\end{figure}

\begin{figure}
\centering
\includegraphics[height=6.5cm, width=9cm]{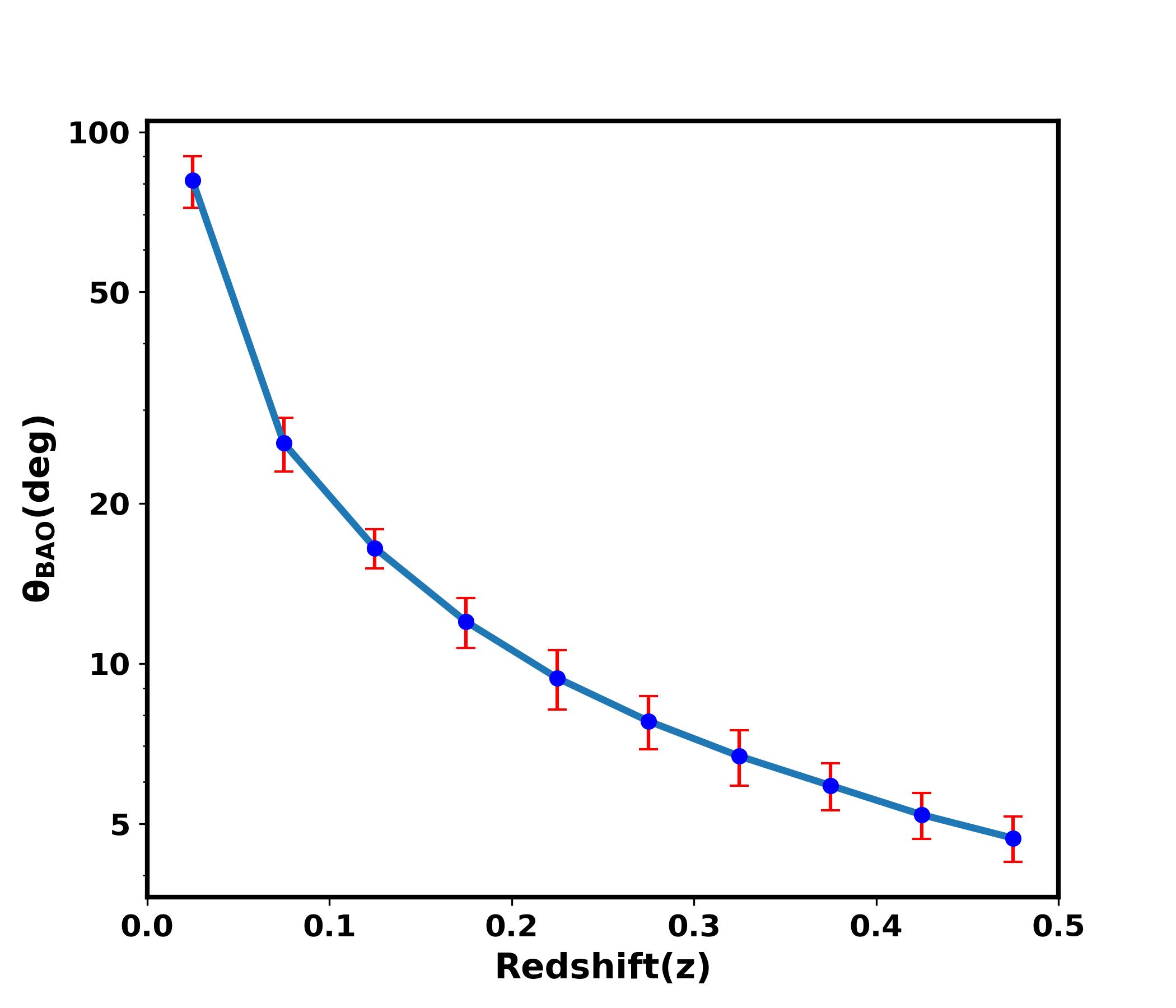}
\caption{We depict the relationship between the BAO scale along with the associated error (in red), primarily arising from photo-z inaccuracies, as a function of redshift (z) up to z = 0.5. This presentation offers crucial insights into how the BAO scale and its error vary with z, highlighting a consistent relative error across measurements. For visual clarity, the error values on the plot have been magnified by a factor of 2.}
\label{fig:baoplot}
\vspace{-0.1cm}
\end{figure}

The covariance of $w(\theta)$, denoted as $Cov_{\theta\theta'}$, is modelled as 
\begin{equation}
    \mathrm{Cov_{\theta\theta'}=\frac{2}{f_{sky}}\sum_{l\geq 0}\frac{2l+1}{(4\pi)^2}P_l(cos(\theta)) P_l(cos(\theta'))\biggl(C_l+\frac{1}{n}\biggr)^2},
\end{equation}
Here, $1/n$ is the shot noise which is related to the number density of galaxies $n$ or, more precisely, the number of objects per steradian, and $f_{sky}$ represents the fraction of the sky that is covered by the survey or observation \citep{cabre2007error, dodelson2020modern}. Figure \ref{fig:covmatplot} illustrates how the covariance matrix, derived from the linear matter power spectrum, changes concerning angular separations at a specific redshift. The increase in the covariance matrix as both $\theta$ and $\theta'$ rise suggests a correlation between these angular scales. Figure \ref{fig:baoplot} provides a comprehensive view of how the BAO scale and its error (calculated by using Equation \ref{eq:fit}) evolve with redshift, emphasizing the impact of photo-z inaccuracies. The consistent relative error observed across measurements serves as a key finding, shedding light on the reliability of BAO scale determinations in the given redshift range. In future work on the application of this technique to data, one can estimate the covariance matrix from simulations.   

Several detectors are set to measure the BAO scale with impressive accuracy, delving into the depths of the universe's redshift. Among these detectors, DESI \citep{collaboration2023early} stands out as a groundbreaking 5-year experiment on the ground, specifically designed to investigate BAO and the changes in cosmic structures as seen through redshift-space distortions. Covering a vast area of 14,000 square degrees in the sky ($f_{\text{sky}} = 0.3$), DESI aims to achieve extremely precise measurements of the BAO feature. The goal is to surpass an impressive precision of 0.5\%, focusing on redshift intervals $0.0 < z < 1.1$, $1.1 < z < 1.9$, and $1.9 < z < 3.7$ \citep{collaboration2023early}. Additionally, Euclid \citep{laureijs2011euclid} and the Vera Rubin Observatory \citep{abell2009lsst, ivezic2019lsst}, covering a vast sky area of 18,000 square degrees will be able to make precise measurements of BAO up to high redshift.

\section{Forecast for reconstructing $\mathcal{F}$ with redshift}
\label{sec:Forecast}

\subsection{Case with the fixed Hubble constant $H_0$}\label{sec:fzfh0}
The intricate connection between the EM wave luminosity distance at a specific redshift (z) and key cosmological parameters, including the BAO scale ($\theta_{BAO}$) and the sound horizon ($r_s$), is succinctly captured by the equation
\begin{equation}
    \mathcal{F}(z)=\frac{d_l^{\rm GW}(z)\theta_{\rm BAO}(z)}{(1+z)r_s}.
    \label{eq:fzeqn}
\end{equation}
This equation serves as a focal point for the frictional component under consideration. In the context of ($\mathcal{F}(z)$), it establishes a connection with three distinct length scales
\texttt{(i)GW luminosity distance:}  The luminosity distance for detectable GW events which is determined using \texttt{Bilby}.
\texttt{(ii) BAO scale($\theta_{BAO}$):} We utilize the BAO scale, derived from the 2PACF using the Equation \ref{eq:fit}. This method is not bound to any particular survey, affording us the flexibility to apply our methodology across a range of surveys without being restricted to a specific one.
\texttt{(iii) Sound Horizons distance:}The sound horizon is measured by analyzing acoustic oscillations in the CMB radiation using several missions such WMAP \citep{hinshaw2013nine}, Planck \citep{aghanim2020planck}, ACTPol \citep{thornton2016atacama}, SPT-3G \citep{sobrin2022design}, and CMB-S4 \citep{abazajian2016cmb}. In particular, the first peak observed in the angular power spectrum of the CMB corresponds to the scale of the sound horizon during the process of recombination. The measured sound horizon value at the drag epoch is found to be $147.09\pm0.26$ Mpc, indicating a high level of precision from Planck \citep{aghanim2020planck}.

 We utilize a Hierarchical Bayesian framework to calculate the posterior distribution of $\mathcal{F}(z)$ as a function of redshift for a total of $n_{GW}$ GW sources. The equation representing the posterior distribution on $\mathcal{F}$(z) for bright sirens is given as \citep{mukherjee2021testing}
\begin{equation}
\begin{aligned}
    &P(\mathcal{F}(z)|\{d^{GW}\},\{z\}) \propto  \Pi(\mathcal{F}(z))\prod_{i=1}^{n_{GW}}\iiint dr_sP(r_s) dd_l^{GW^i} \\
    \times\, &d\theta_{BAO}^i P(\theta_{BAO}^i) \mathcal{L}(d_l^{GW^i}|\mathcal{F}(z^i),\theta_{BAO}^i,r_s,z^i).
\end{aligned}
\label{eq:FZposterior}
\end{equation}
In this equation, $\mathrm{P(\mathcal{F}(z)|\{d^{GW}\}, \{z\})}$ denotes the posterior probability density function of $\mathrm{\mathcal{F}(z)}$ given the GW source data $d^{GW}$ and spectroscopic redshift from EM counterpart of these sources $\{z\}$ for $i\in n_{GW}$ events. The measurements of BAO angular scale $\theta_{BAO}$ at the redshift (z) of the GW source where an EM counterpart of GW source can be measured using the galaxy survey, and it is marginalized over. The term $\mathcal{L}(d_l^{GW^i}|\mathcal{F}(z)^i, \theta_{BAO}^i, r_s, z^i)$ corresponds to the likelihood function. $P(r_s)$ and $P(\theta_{BAO}^i)$ denote the prior probability distributions for the sound horizon $(r_s)$ and BAO scale respectively. The term $\Pi(\mathcal{F}(z))$ represents the prior distribution for $\mathcal{F}(z)$, which encapsulates our knowledge or assumptions about the frictional term before incorporating the measured data. We adopt a flat prior on the frictional term, $\mathcal{F}(z)$ from 0.1 to 2.
For this investigation, we make specific assumptions regarding the probability distribution of redshift(z) and prior to the frictional term $\mathcal{F}(z)$. In this analysis, we consider bright sirens only. So, we assume that the host galaxy redshift of the GW sources is measured spectroscopically. In this case, it implies that we have precise knowledge or assume that the redshift of the sources is concentrated at a particular value. 

\begin{figure*}
    \centering
    \includegraphics[height=11.3cm, width=18.7cm]{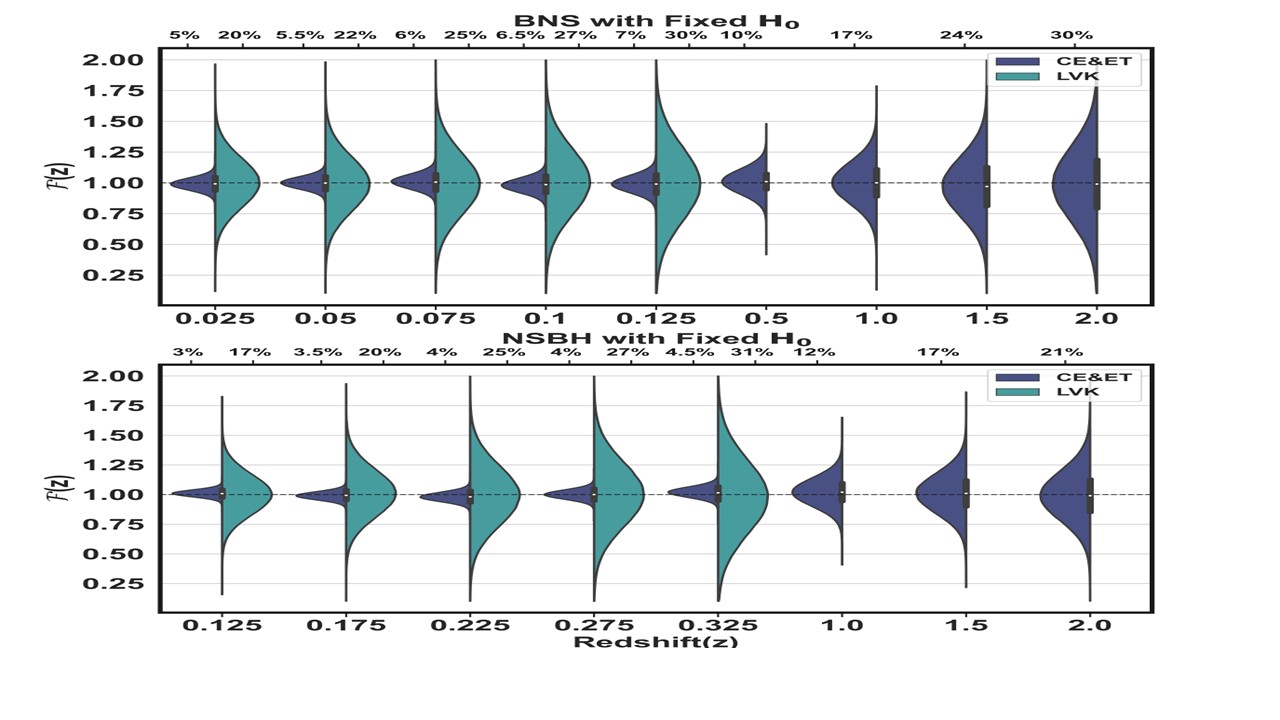}
    \caption{
    This violin plot illustrates the posterior on the reconstruction of the non-GR parameter $\mathcal{F}(z)$ as a function of cosmic redshift is feasible from LVK (in green) and CE \& ET (in blue). The top and bottom plot relates to possible measurement using one BNS  and NSBH source respectively, both assuming a fixed Hubble constant. The upper x-axis denotes the associated $1$-$\sigma$ error bars for each measurement. It demonstrates the potential to measure $\mathcal{F}(z)$ with greater precision and at significantly deeper cosmic redshifts using future detectors such as CE \& ET, as opposed to the limitations of our current LVK detector.}
    \label{fig:violinplot1}
\end{figure*}

In Figure \ref{fig:violinplot1}, we presented the measurement of $\mathcal{F}(z)$, with a fixed value of $H_0$ for both the CE \& ET and the LVK system, focusing on a single event. This plot illustrates a notable trend where the measurement error increases with higher redshifts. Additionally, it discerns that, at the same redshift, the error on $\mathcal{F}(z)$ is more pronounced for BNS systems compared to NSBH systems, attributed to the lower SNR observed for BNS events.

\subsection{Case with varying the Hubble constant $H_0$}\label{sec:fzvh0}

In the preceding sections, our analysis primarily focused on evaluating the frictional term denoted as $\mathcal{F}(z)$ in Equation \ref{eq:fzeqn}, assuming a fixed Hubble constant ($H_0$). However, in this section, we jointly infer both $\mathcal{F}(z)$ and $H_0$. This extension enables the simultaneous estimation of both the frictional term $\mathcal{F}(z)$ and $H_0$, derived from sound horizon measurements that remain constant across redshift, serving as an overall normalization on $\mathcal{F}(z)$ measurements.

In this study, our primary focus is to constrain the frictional term $\mathcal{F}(z)$. Simultaneously, while measuring this term, we also estimate the value of the $H_0$. Leveraging the capabilities of our future planned detectors, we can extend our observations to deeper redshifts, allowing us to measure $H_0$ either in conjunction with $\mathcal{F}(z)$ or independently. This innovative approach holds the potential to address the Hubble tension\citep{Abdalla:2022yfr}, an observed discrepancy between the locally measured value of the $H_0$ \citep{Riess:2021jrx} and the value inferred from CMB observations \citep{Planck:2018vyg}.

Employing a hierarchical Bayesian analysis, our unified framework efficiently estimates these two parameters by incorporating diverse observational datasets, including GW measurements, CMB measurements, and large-scale structure surveys, as discussed in previous sections. The hierarchical organization of these datasets ensures robust integration, effectively handling their unique uncertainties and systematics. The resulting correlated constraints significantly contribute to our understanding of cosmological phenomena, addressing potential discrepancies between datasets and revealing insightful connections among fundamental cosmological parameters.

The equation representing the joint posterior distribution on $\mathcal{F}$(z) and $H_0$ for these sources is given as \citep{mukherjee2021testing}

\begin{equation}
\begin{aligned}
    &P(\mathcal{F}(z),H_0|\{d^{GW}\},\{z\}) \propto  \Pi(\mathcal{F}(z))\Pi(H_0)\prod_{i=1}^{n_{GW}}\iiint  \\ \times\, & dr_sP(r_s^i) dd_l^{GW^i} 
    d\theta_{BAO}^i  P(\theta_{BAO}^i) \mathcal{L}(d_l^{GW^i}|\mathcal{F}(z^i),H_0,\theta_{BAO}^i,r_s,z^i).
\end{aligned}
\label{FZ}
\end{equation}
In this equation, $\mathrm{P(\mathcal{F}(z), H_0|\{d^{GW}\}, \{z\})}$ denotes the joint posterior probability density Function of $\mathrm{\mathcal{F}(z)}$ and $H_0$. The term $\mathcal{L}(d_l^{GW^i}|\mathcal{F}(z)^i, H_0^i, \theta_{BAO}^i, r_s,z^i)$ corresponds to the likelihood function. The term $\Pi(H_0)$ denotes the prior distribution for $H_0$, encapsulating our knowledge or assumptions regarding the Hubble constant before incorporating measured data. All other terms represent the same entities as defined in Section \ref{sec:fzfh0}. In our hierarchical Bayesian framework, we adopt a flat prior assumption, indicating a lack of prior knowledge about the parameters for both the frictional term ($\mathcal{F}(z)$) and the cosmic expansion rate ($H_0$). The parameter ranges are defined from 0.1 to 2.0 for $\mathcal{F}(z)$ and from 40 km/s/Mpc to 100 km/s/Mpc for $H_0$.

\begin{figure*}
    \centering
    \includegraphics[height=11.2cm, width=18.7cm]{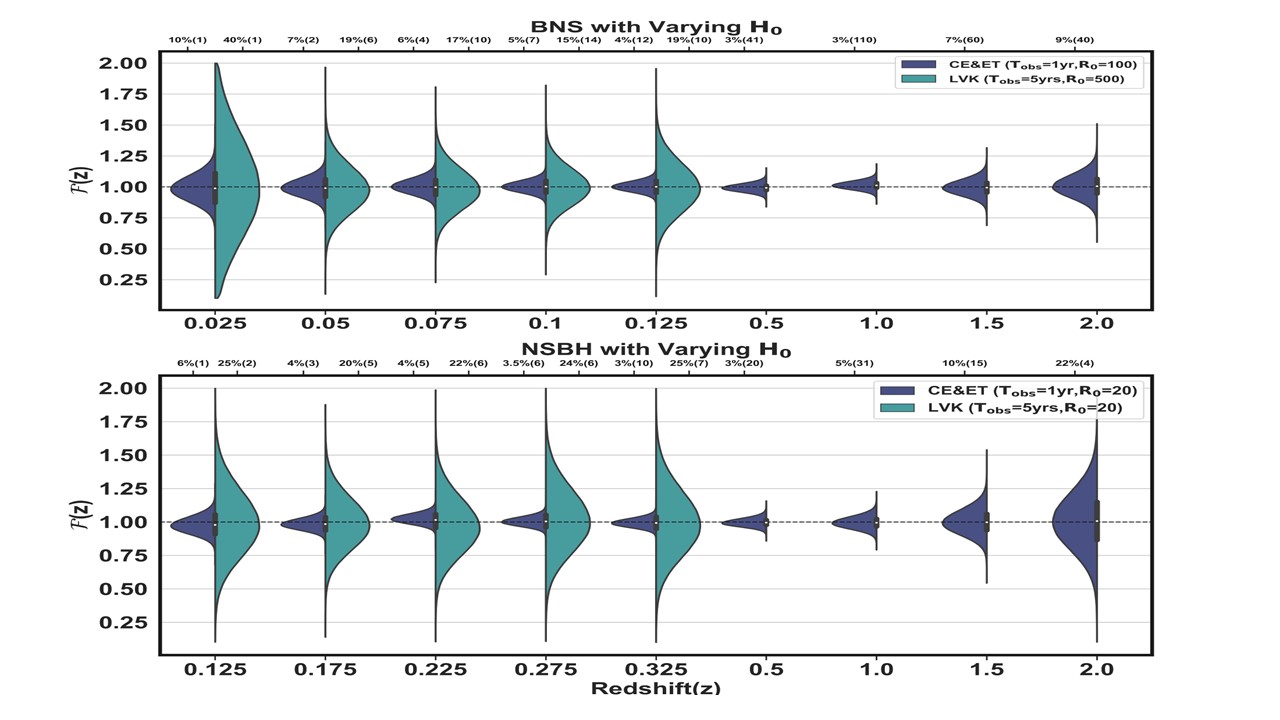}
    \caption{
    This violin plot illustrates the posterior on the reconstruction of the non-GR parameter $\mathcal{F}(z)$ as a function of cosmic redshift is feasible from LVK (in green) and CE \& ET (in blue). The top and bottom plot relates to possible measurement using one BNS and NSBH source respectively jointly with the Hubble constant. The upper x-axis denotes the associated $1$-$\sigma$ error bars for each measurement, with the number in parentheses representing the total number of events in that specific redshift bin.
    The plot highlights the potential for more accurate measurements of $\mathcal{F}(z)$ at significantly deeper cosmic redshifts, particularly with future detectors like CE \& ET, in contrast to the limitations posed by our current LVK detector, similar to the case with fixed Hubble constant. Additionally, it underscores that we can achieve more precise measurements of $\mathcal{F}(z)$ for BNS systems compared to NSBH events, given the higher likelihood of BNS events occurring.}
    \label{fig:violinplot2}
\end{figure*}

In Figure \ref{fig:violinplot2}, we present measurements of $\mathcal{F}(z)$ with varying values of $H_0$ for both the CE \& ET and the LVK system, focusing on the total number of events within each redshift bin (mentioned in the parenthesis in the upper x-axis in the plot). This plot reveals a discernible trend, indicating an increase in measurement errors with higher redshifts. Notably, due to the higher likelihood of BNS events compared to NSBH events, we can achieve a more accurate measurement of $\mathcal{F}(z)$ for BNS events than for NSBH events.

\section{Result}
\label{sec:result}

\subsection{Reconstruction of $\mathcal{F}(z)$}\label{sec:recfz}
This study exclusively focuses on bright sirens, indicating NSBH and BNS mergers with distinct EM counterparts \citep{nakar2018gamma, kasliwal2017illuminating, arcavi2018first, troja2017x, hallinan2017radio, ghirlanda2019compact, smartt2017kilonova, kasen2017origin, janka1999black, foucart2012black, foucart2018remnant}. These include short GRBs and high-energy gamma-ray bursts detected by satellites like Fermi \citep{bissaldi2009ground} and Swift \citep{burrows2005swift}. Additionally, kilonovae generate bright optical and infrared transients, offering insights into heavy element formation. Observatories like the Vera Rubin \citep{abell2009lsst}, HST \citep{scoville2007cosmos}, and the upcoming Roman space telescope \citep{eifler2021cosmology} are poised to capture these events. Post-merger, afterglows spanning X-ray, UV, optical, and radio wavelengths are scrutinized with instruments such as Chandra \citep{vikhlinin2009chandra}, VLA \cite{benz1989vla}, ZTF \citep{bellm2014zwicky}, Astrosat \citep{agrawal2006broad}, Daksha \citep{bhalerao2022science} and GMRT \citep{bhattacharyya2016gmrt}.

Detecting EM counterparts in BNS and NSBH mergers poses challenges related to prompt collapse, ejecta mass, and orientation. The visibility of collimated jets is influenced by their opening angle, with wider angles increasing detectability. Distance, instrument sensitivity, and rapid event localization further complicate detection. In astrophysics, terms like "structured jets" and "off-axis viewing" describe these phenomena \citep{rees1998refreshed, ramirez2001quiescent, kumar2000some}. Coordinated efforts between GW and EM observatories are vital for insights into these cosmic collisions. Despite progress, challenges persist, requiring proximity, orientation, and sensitivity considerations. Success relies on refining theoretical models, combining GW and EM observations, and technological advancements. Future endeavors aim to optimize multi-messenger observations, deepen theoretical comprehension, and amplify the detectability of NSBH merger EM counterparts.

In our study, we present cases for both the CE \& ET and LVK systems. Specifically for CE \& ET, we showcase outcomes for up to redshifts $z =2$ for different numbers of detectable events. In the case of NSBH binaries, we consider a total of 30 events with a local merger rate ($R_0$) set at 20 $\rm Gpc^{-3}yr^{-1}$ in the LVK system. These events span redshifts from 0.125 to 0.425. Similarly, for BNS systems, we analyze approximately 25, 40, and 60 events with $R_0$ set to 200 $\rm Gpc^{-3}yr^{-1}$, 300 $\rm Gpc^{-3}yr^{-1}$, and 500 $\rm Gpc^{-3}yr^{-1}$, respectively. The redshifts of these events range from 0.025 to 0.225 over a 5-year observation period. In Figure \ref{fig:BNSevent} and  Figure \ref{fig:NSBhevent}, the number of detectable events as a function of redshift with different $R_0$ and $T_{obs}$ are presented for the LVK and CE \& ET systems. 

\begin{figure*}
\centering
\includegraphics[width=height=7.0cm, width=18.7cm]{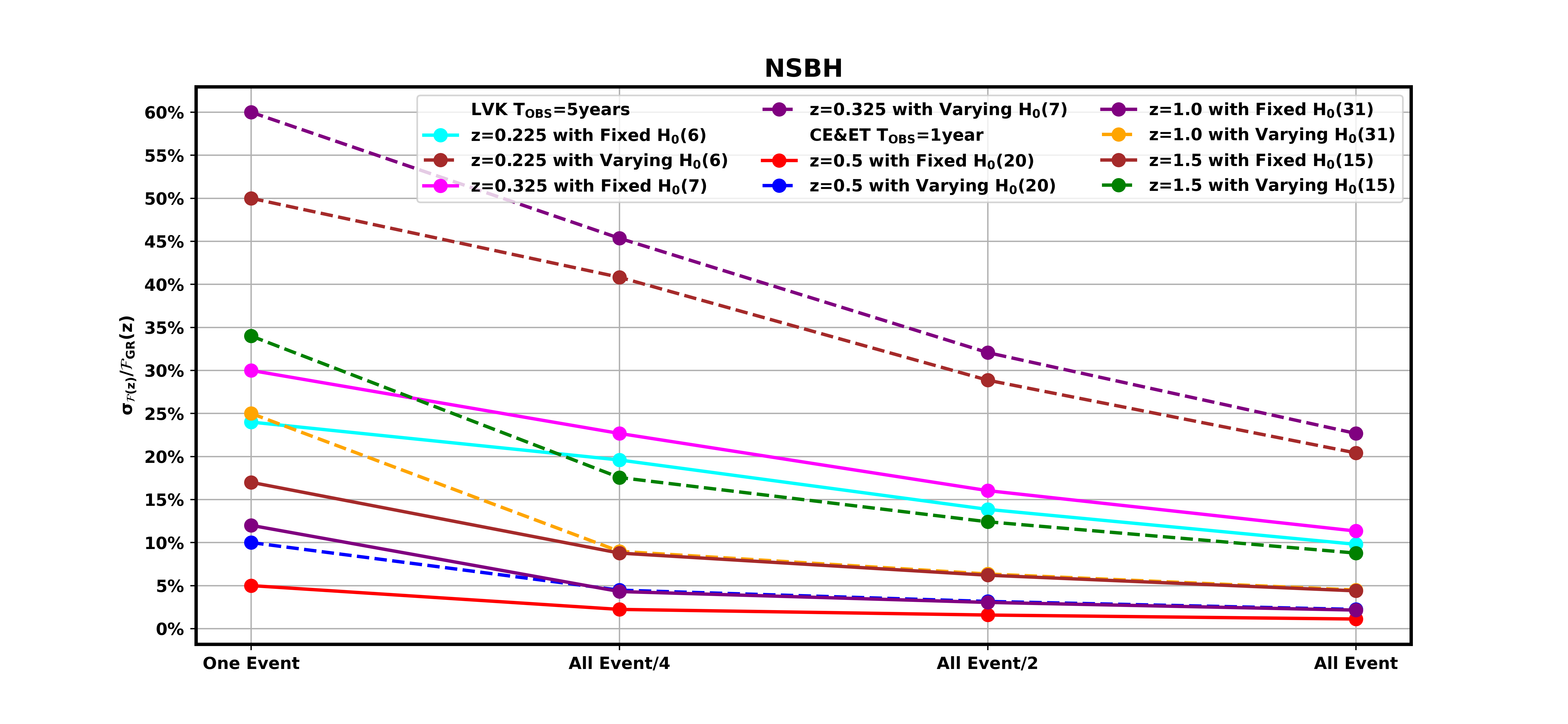}
\caption{We show the $1$-$\sigma$ error-bar in the measurements of $\mathcal{F}(z)$ as a function of redshift and the number of GW events for NSBH events. The comparison is made between scenarios with a constant Hubble constant ($H_0$) and for the case with $H_0$ inferred jointly with $\mathcal{F}(z)$.  All the events at each redshift bin are mentioned in parentheses in the figure legend.}
\label{fig:NSBH}
\end{figure*}

\begin{figure*}
\centering
\includegraphics[height=7.5cm, width=18.7cm]{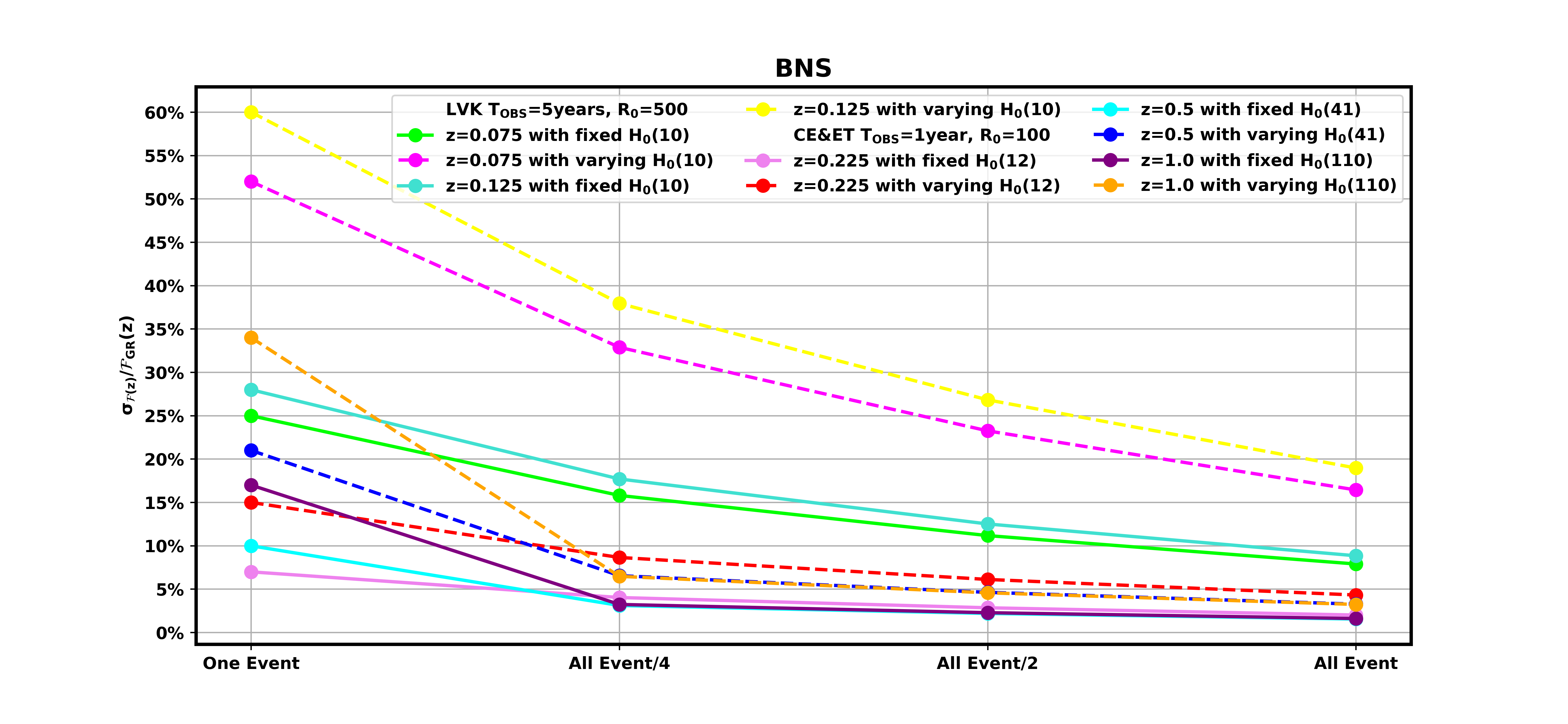}
\caption{We show the $1$-$\sigma$ error-bar in the measurements of $\mathcal{F}(z)$ as a function of redshift and the number of GW events for BNS events. The comparison is made between scenarios with a constant Hubble constant ($H_0$) and those with varying $H_0$.  All the events at each redshift bin are mentioned in parentheses in the figure legend.}
\label{fig:BNS}
\end{figure*}

Figure \ref{fig:NSBH} and Figure \ref{fig:BNS} provide a comprehensive comparison of $\mathcal{F}(z)$ measurements with fixed $H_0$ and varying $H_0$ across different scenarios for both LVK and CE \& ET. These scenarios include the detection of a single event, a quarter of the events, half of the events, and the detection of all events. The aim is to demonstrate how the precision of measurements, ranging from percentages to sub-percentages, can be achieved using future detectors. This comparative analysis sheds light on the impact of event detection rates on the precision of $\mathcal{F}(z)$ measurements and underscores the potential of advanced detectors like CE \& ET to provide increasingly accurate cosmological insights. In this section, we measure $\mathcal{F}(z)$ with a constant $H_0$ that remains uniform across redshifts, serving as an overall normalization factor for $\mathcal{F}(z)$ measurements. Consequently, if an independent precise measurement of $\mathcal{F}(z)$ in our local universe is possible, it will allow us to normalize it to 1, eliminating the dependence on $H_0$. This normalization proves effective particularly for low redshifts, where the cumulative effects of modified gravity wave propagation have not had sufficient time to accumulate deviations from GR.

The error associated with the frictional term approximately scales as the inverse of the square root of the number of GW sources represented as $\mathrm{1/\sqrt{N_{GW}}}$. This implies that as the number of GW sources increases, the error on the frictional term decreases, thereby improving the precision of the measurement. However, this improvement reaches a saturation point, beyond which an increase in the number of GW sources does not lead to a significant enhancement in the measurement's accuracy. This saturation is primarily due to the presence of error on the BAO scale. The BAO scale error acts as a limiting factor in the precision of the frictional term estimation. Therefore, to achieve further improvement in the estimation of the frictional term with redshift, a more precise measurement of the BAO scale is required. In essence, the precision of the frictional term estimation is a delicate balance between the number of GW sources and the accuracy of the BAO scale measurement and GW luminosity distance. Optimizing these factors can lead to significant advancements in our understanding of $\mathcal{F} (z)$ and discovering unexplored territories in fundamental physics. On the GW side, the measurements of $\mathcal{F}(z)$ are solely dependent on $D_l^{GW}(z)$. Any improvement in the precision of $D_l^{GW}(z)$ will markedly enhance the accuracy of $\mathcal{F}(z),$ as $D_l^{GW}(z)$ stands out as the predominant source of error for $\mathcal{F}(z)$. Given that $D_l^{GW}(z)$ is degenerate with other GW parameters, most notably the inclination angle ($i$), measuring the inclination angle from alternative sources (such as EM counterparts) can substantially reduce the error in $D_l^{GW}(z)$ and consequently enhance the precision of $\mathcal{F}(z)$ \citep{xie2023breaking}. For redshift inference from EM counterparts, the influence of peculiar velocity contamination is particularly significant at low redshifts ($z<0.03$) but it is not important for high redshift \citep{Mukherjee:2019qmm, nimonkar2024dependence}. As most of the detected sources are at high redshift, we have ignored this effect. Also, weak lensing of the GW sources can be a source of contamination to the luminosity distance up to a few percent at redshifts above $z=0.5$ \citep{2010PhRvD..81l4046H}. In comparison to the error bar on luminosity distance, the weak lensing error bar is small. As most of the sources with better luminosity distance measurement are mainly at low redshift ($z<0.5$), we are getting maximum constraints around this redshift. So, the weak lensing uncertainty is not a major contamination for these sources. Moreover, weak lensing can provide additional information to constraint non-GR parameter space \citep{mukherjee2020multimessenger, Mukherjee:2019wcg, Mpetha:2022xqo, balaudo2023prospects}, which we have not considered in this analysis.

\subsection{Reconstruction of $\mathcal{F}(z)$ and $H_0$}
\label{sec:ReconofH0andFz}

By extension of the hierarchical Bayesian framework used in the previous section to measure $\mathcal{F}(z)$ with fixed $H_0$, we can jointly measure $\mathcal{F}(z)$ and $H_0$. In our framework, we measure the EM luminosity distance from different length scales, namely the BAO scale, sound horizon distance, and redshift. In Figure \ref{fig:NSBH} and Figure \ref{fig:BNS}, we offer a thorough comparison of $\mathcal{F}(z)$ measurements with both fixed and varying $H_0$ across different scenarios for both the LVK and CE \& ET systems. 
It is noteworthy that when allowing for variations in $H_0$, the errors in the measurements of $\mathcal{F}(z)$ increased by a factor ranging from 2 to 2.2 compared to measurements with a fixed $H_0$. In Figure \ref{fig:fzasz}, we highlight the enhanced precision in measuring the redshift-dependent variation of the Planck mass ($\mathcal{F}(z)$) for NSBH and BNS systems. The figure emphasizes that the measurement of $\mathcal{F}(z)$ is notably improved in the redshift range of 0.2 to 1.0. The precision of estimating $\mathcal{F}(z)$ depends on accurate measurements of both the BAO scale and $D_l^{GW}(z)$. Improved accuracy in the BAO scale measurements directly enhances precision in estimating $\mathcal{F}(z)$. On the GW side, accurate measurements of $\mathcal{F}(z)$ rely solely on $D_l^{GW}(z)$. Enhancing the precision of $D_l^{GW}(z)$ improves $\mathcal{F}(z)$, as mentioned in the previous Subsection \ref{sec:recfz}. There can also be peculiar velocity corrections and weak lensing contamination at low redshift as described in \ref{sec:recfz}  which are not taken care of in this study.

\section{conclusion and discussion}
\label{sec:conclusion}
In this study, we present a model-independent approach to search for the frictional term, which can capture deviations at any redshift from GR. We propose a model-independent reconstruction of the frictional term as a function of redshift using a data-driven approach by amalgamating three length scales with redshift namely the luminosity distance from GW sources $d_l^{\rm GW}(z)$, the angular scale of BAO scale from galaxy surveys $\theta_{\rm BAO} (z)$, and the sound horizon $r_s$ from CMB observations. We show this provides a unique opportunity to scrutinize theories of gravity, specifically to test the presence of a frictional term. In the literature, various model-dependent searches for the frictional term have been conducted. These searches face challenges, especially near redshift z=0 and at large redshifts. Particularly noteworthy are the difficulties encountered when there is a potential loss of gravitons in extra dimensions. Furthermore, in the intermediate redshift range, the frictional term may display oscillatory behavior that cannot be adequately captured by these model-dependent searches. 

We demonstrate that by integrating BAO measurements from galaxies, sound horizon distances from the CMB power spectrum, redshift information from the EM counterpart, and luminosity distance measurements from GW sources, one can perform a comprehensive measurement of cosmological parameters. Specifically, we highlight the potential to jointly determine the Hubble constant ($H_0$) and the frictional term through this integrated analysis. We have demonstrated the feasibility of measuring the frictional term using our ongoing detectors, such as LVK, across various redshifts. Additionally, we have explored the capabilities of upcoming detectors, specifically CE \& ET, in quantifying the frictional term under various scenarios. Our analysis includes cases where a fraction of the event's EM counterpart is detectable, acknowledging the inherent difficulty in detecting EM counterparts.

 We demonstrate the feasibility of achieving a precision measurement of the frictional term, ranging from a percent to sub-percent accuracy, as a function of cosmic redshift. This is achievable using data from LVK and, in the future, from CE \& ET, despite the limitation of a finite number of bright sirens. The measurement of $\mathcal{F}(z)$ is particularly improved in the redshift range of 0.2 to 1.0. This improvement is attributed to two main factors: SNR and the number of events. The measurements from LVK can be further improved by including LIGO-India \citep{saleem2021science}\footnote{\url{https://dcc.ligo.org/LIGO-M1100296/public}}.

 In the ($\Xi_0$, n) parametrizations of the frictional term, the fiducial value for $\Xi_0$ is set to 1. However, it is important to note that the parameter n is not defined at this fiducial value of $\Xi_0$=1. Consequently, measuring the value of n becomes increasingly challenging as $\Xi_0$ approaches 1 \citep{belgacem2018modified}. By utilizing 1000 bright sirens up to redshift 8, \citet{belgacem2018modified} demonstrated that $\Xi_0$ can be measured with an accuracy of 0.8\% using the Einstein Telescope for a fixed value of n. In our study, employing a model-independent parametrization, we found that with 1 year of observation using CE \& ET, we can measure the frictional term with an accuracy of 1.62\% and 2\% for BNS and NSBH systems at redshift 0.5. Additionally, we can reconstruct the frictional term as a function of redshift without relying on any specific parametrization, as illustrated in Figure \ref{fig:fzasz}. This error scales inversely with the square root of the observation time, i.e., as $1/\sqrt{T_{obs}/years}$.
 
Our proposed technique extends to dark sirens, where redshift inference becomes necessary. In such cases, we leverage the three-dimensional clustering scale of GW sources with galaxies to infer redshift information \citep{mukherjee2018beyond, mukherjee2020probing, abbott2021constraints}. Furthermore, this methodology holds promise for both bright and dark standard sirens in the mHz range, detectable by future space-based detectors like LISA. In this study, our exclusive focus is on bright sirens, and as a result, we exclude a prominent category of GW sources Binary Black Holes (BBH) due to their unlikely association with EM counterparts. Additionally, there are challenges associated with detecting EM counterparts for BNS and NSBH systems \citep{chase2022kilonova, Ronchini:2022gwk}. One potential method to enhance the accuracy of frictional term measurements is to include dark sirens, which lack EM counterparts. In the future, the integration of LIGO-India and the LISA GW detector, in conjunction with projects such as eBOSS, DESI, Vera Rubin, and Euclid for BAO measurements, along with the application of this technique to dark sirens, holds the potential to substantially improve the precision of measuring the frictional term.

\section*{Acknowledgements}
The authors express their gratitude to Guillaume Dideron for reviewing the manuscript and providing useful comments as a part of the LIGO publication policy.
This work is part of the \texttt{⟨data|theory⟩ Universe-Lab}, supported by TIFR and the Department of Atomic Energy, Government of India. The authors express gratitude to the computer cluster of \texttt{⟨data|theory⟩ Universe-Lab} and the TIFR computer center HPC facility for computing resources. Special thanks to the LIGO-Virgo-KAGRA Scientific Collaboration for providing noise curves. LIGO, funded by the U.S. National Science Foundation (NSF), and Virgo, supported by the French CNRS, Italian INFN, and Dutch Nikhef, along with contributions from Polish and Hungarian institutes. This collaborative effort is backed by the NSF’s LIGO Laboratory, a major facility fully funded by the National Science Foundation. The research leverages data and software from the Gravitational Wave Open Science Center, a service provided by LIGO Laboratory, the LIGO Scientific Collaboration, Virgo Collaboration, and KAGRA. Advanced LIGO's construction and operation receive support from STFC of the UK, Max-Planck Society (MPS), and the State of Niedersachsen/Germany, with additional backing from the Australian Research Council. Virgo, affiliated with the European Gravitational Observatory (EGO), secures funding through contributions from various European institutions. Meanwhile, KAGRA's construction and operation are funded by MEXT, JSPS, NRF, MSIT, AS, and MoST.

 This material is based upon work supported by NSF’s LIGO Laboratory which is a major facility fully
funded by the National Science Foundation. We acknowledge the use
of the following packages in this work: Astropy \citep{robitaille2013astropy, price2018astropy}, Bilby \citep{ashton2019bilby}, Pandas \citep{mckinney2011pandas}, NumPy \citep{harris2020array}, Seaborn \citep{bisong2019matplotlib}, CAMB \citep{lewis2011camb}, Scipy \citep{virtanen2020scipy}, Matplotlib \citep{4160265}, Dynesty \citep{speagle2020dynesty}, and emcee \citep{foreman2013emcee}

\section*{Data Availability}
The corresponding author will provide the underlying data for this article upon request.

\bibliographystyle{mnras}
\bibliography{references} 
\label{lastpage}
\appendix
\section{Uncertainties in $H_0$ in the Reconstruction of $H_0$ and $\mathcal{F}(z)$}
\label{sec:appendix}

In Section \ref{sec:ReconofH0andFz}, we extended the hierarchical Bayesian framework previously employed for measuring $\mathcal{F}(z)$ with a fixed $H_0$ to a joint measurement scenario for both $\mathcal{F}(z)$ and $H_0$. Within this framework, a flat prior assumption was utilized, indicating a lack of prior knowledge about the parameters associated with the frictional term ($\mathcal{F}(z)$) and the cosmic expansion rate ($H_0$). For the parameters governing the frictional term ($\mathcal{F}(z)$) and the Hubble constant ($H_0$), we established parameter ranges of 0.1 to 2.0 and 40 km/s/Mpc to 100 km/s/Mpc, respectively. 

\begin{figure}
\centering
\includegraphics[height=6.2cm, width=9cm]{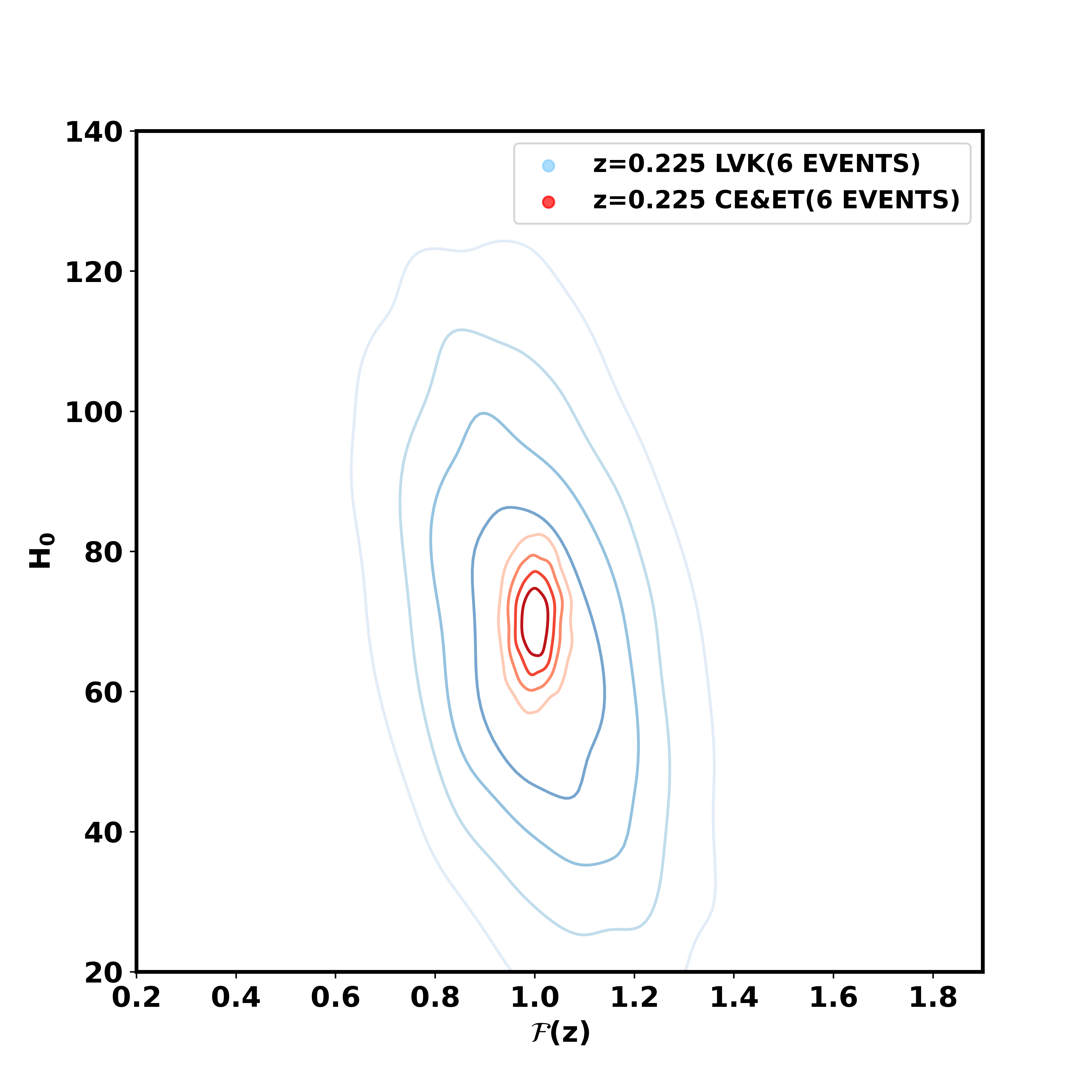}
\caption{In this plot, we depict the joint posterior of $\mathcal{F}(z)$ and $H_0$ for CE \& ET and LVK for the NSBH events for a total number of events scenario at z=0.225. In the parenthesis, it depicts the total no of events at that particular redshift.}
\label{fig:contour}
\vspace{-0.5cm}
\end{figure}

\begin{figure}
\centering
\includegraphics[height=6.2cm, width=9cm]{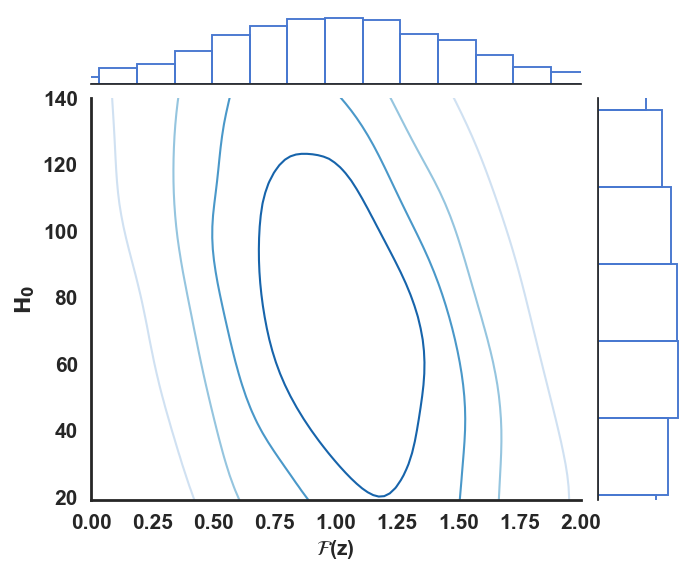}
\caption{In this plot,we depict the Joint Posterior of $\mathcal{F}(z)$ and $H_0$ upto 4 $\sigma$ level for LVK for the NSBH events scenario at z=0.225 for one event.}
\label{fig:FzH0LVK}
\vspace{-0.5cm}
\end{figure}

\begin{figure}
\centering
\includegraphics[height=6.2cm, width=9cm]{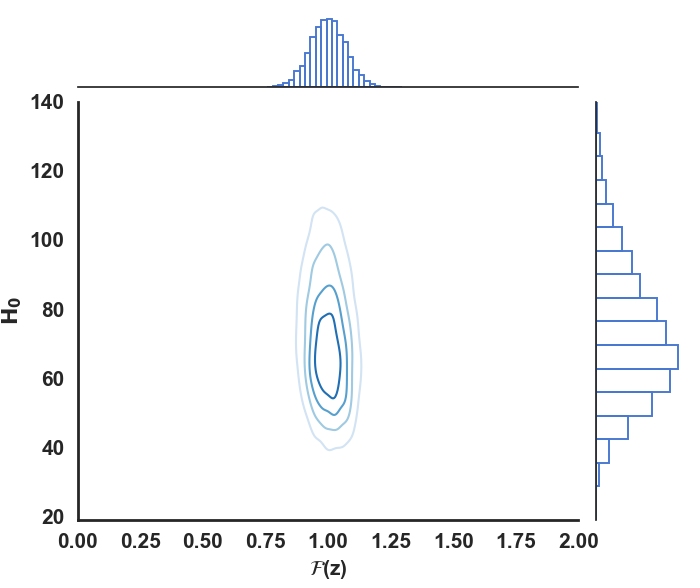}
\caption{In this plot, we depict the joint posterior of $\mathcal{F}(z)$ and $H_0$ upto 4 $\sigma$ level for CE \& ET for the NSBH events scenario at z=0.225 for one event.}
\label{fig:FzH0CEET}
\vspace{-0.5cm}
\end{figure}

In  Figure \ref{fig:contour}, we showcase the joint posterior of $\mathcal{F}(z)$ and $H_0$ for two distinct scenarios. The first scenario corresponds to CE \& ET, considering all observed events for z=0.225 in the case of NSBH events. The second scenario pertains to LVK, also for z=0.225 also focusing on NSBH events. This visualization provides a comparative analysis, illustrating the potential advancements achievable with our future detector, CE \& ET, in contrast to the current capabilities of the LVK detector. The joint posterior distribution offers valuable insights into the precision and capabilities of each detector in estimating the parameters $\mathcal{F}(z)$ and $H_0$. In Figure \ref{fig:FzH0LVK} and Figure \ref{fig:FzH0CEET} we showed the joint posterior of $\mathcal{F}(z)$ and $H_0$ for  CE \& ET and LVK at redshift 0.225 for just only one event.

\bsp
\end{document}